\documentclass[12pt]{article}
\usepackage{amsmath,amssymb}
\usepackage{bm}
\usepackage{color}
\usepackage{graphicx}
\usepackage[dvipdfmx]{epsfig}

\newcommand{\tr}[1]{\hspace{.1em}\mathrm{tr\hspace{.1em}#1}}
\begin{document}
\title{
\begin{flushright}
\ \\*[-80pt] 
\begin{minipage}{0.2\linewidth}
\normalsize
HUPD-1910 \\*[50pt]
\end{minipage}
\end{flushright}
{\Large \bf 
Hidden relations 
in three generation seesaw model \\
with Dirac mass matrix of four-zero texture 
\\*[20pt]}}

\author{ 
\centerline{
Takuya~Morozumi$^{1,2}$\footnote{E-mail address: morozumi@hiroshima-u.ac.jp},
~Yusuke~Shimizu$^{1,}$\footnote{E-mail address: yu-shimizu@hiroshima-u.ac.jp}, 
Hiroyuki~Umeeda$^{3}$\footnote{E-mail address:
umeeda@gate.sinica.edu.tw}
and
~Akihiro~Yuu$^{1,}$\footnote{E-mail address:d162264@hiroshima-u.ac.jp}} \\
\\*[20pt]
\centerline{
\begin{minipage}{\linewidth}
\begin{center}
$^1${\it \normalsize
Graduate~School~of~Science,~Hiroshima~University, \\
Higashi-Hiroshima~739-8526,~Japan} \\*[5pt]
$^2${\it \normalsize 
Core~of~Research~for~the~Energetic~Universe,~Hiroshima~University, \\
Higashi-Hiroshima~739-8526,~Japan} \\*[5pt]
$^3${\it \normalsize 
Institute of Physics, Academia Sinica,
\\
Nangang, Taipei 11529, Taiwan} \\*[5pt]
\end{center}
\end{minipage}}
\\*[50pt]}

\date{
\centerline{\small \bf Abstract}
\begin{minipage}{0.9\linewidth}
\medskip 
\medskip 
\small
We present predictions for CP violating phases of the Type-I seesaw model with four-zero textures on the Dirac mass matrix. For the four-zero textures, the effective low energy Majorana mass
matrix is parametrized with seven parameters. They are three mass-dimensional parameters, two angles and two CP violating sources. The number of these parameters is less than that of the general
description of the Majorana mass matrix with
three neutrino masses, three mixing angles and three CP violating phases.  In particular, only two independent  CP violating sources give rise to three CP violating phases.
The efficient and comprehensive method is proposed in this paper to investigate four-zero textures in Type-I seesaw. 
We numerically show the possible range of CP violating phases in the plane of a Dirac CP violating phase and one of Majorana phases. 
Some cases show the strong correlations among two phases. 
These correlations can be explained by hidden relations among the elements of Majorana matrix with the four-zero textures.
The hidden relations  are classified according to the position of one vanishing off-diagonal element of Majorana mass matrix.  
The Majorana mass matrix all of whose elements are non-vanishing also produces other hidden relations particularly in the case of four-zero textures.   
By applying the hidden relations, we describe the concrete correlations among CP violating phases.
\end{minipage}
}

\begin{titlepage}
\maketitle 
\thispagestyle{empty}
\end{titlepage}

\section{Introduction}
The advances of the neutrino oscillation experiment reveal 
the many interesting results~\cite{Fukuda:1998mi}-\cite{NuFIT}.
Three mixing angles, two mass squared differences and constraints on Dirac CP violating phases are obtained.
In the future, one may obtain the effective mass for neutrino-less double beta decay.  
It has not been known yet, how these measurements uncover the origin of neutrino mass.
Type-I seesaw model with three right-handed neutrinos is one of its promising candidates~\cite{Minkowski:1977sc}. 
This mechanism leads to effective Majorana mass term for light active neutrinos.
In the model, the effective Majorana mass matrix  is written with Dirac mass matrix and Majorana mass matrix for right-handed neutrinos.
In the real diagonal basis for the Majorana mass matrix and charged-lepton mass matrix, 
the Dirac mass matrix is parametrized by 15 parameters. By adding three real diagonal masses for heavy Majorana neutrinos, the effective Majorana  mass matrix 
is written in terms of 18 parameters. 
  There are many studies to determine the parameters of seesaw model using  
measurements in low energy experiments.  However, as mentioned above, we have more parameters than the number of the measurements.

Some flavor symmetries \cite{Adhikary:2009kz, Adhikary:2012kb, Asai:2018ocx} restrict the forms of the Dirac mass matrix and the Majorana mass matrix.
The models with flavor symmetry have less parameters, which can be determined with the measurements.
In the approach  based on zero textures of the Dirac mass matrix,   
one can also reduce the number of parameters by assigning  vanishing  elements in Dirac mass matrix. The position of zeros can be restricted by some flavor symmetries. 
In this paper, we focus on the four-zero texture model \cite{Choubey:2008tb, Adhikary:2009kz, Adhikary:2012kb, Branco:2007nb} for the Dirac mass matrix since the effective Majorana mass
matrix can be parametrized with seven parameters, which are comparable to the number of measurements.

 Within the four-zero texture model, 
there are still
many possible forms for $3 \times 3$ Dirac neutrino mass matrix.  Indeed,
we have $_9 C_4=126$ different  configurations of  Dirac mass matrix.
 We  develop an efficient method to explore all the configurations without examining each form independently.
In the method, the Dirac matrices related to each other  by permutation of their rows and columns are classified into several 
groups.  Phenomenological constraints on the parameters are also imposed  according to this classification.  
 
The effective Majorana matrix has the expression in terms of nine phenomenological parameters.  Six of them are  three masses and three mixing angles~\cite{Pontecorvo:1957qd,Maki:1962mu} and the others correspond to CP violating phases
(a Dirac CP violating phase~\cite{Kobayashi:1973fv} and two Majorana phases~\cite{Doi:1980yb}).
Since the effective mass matrix in the four-zero texture model is parametrized in terms of seven parameters (five real parameters and two CP violating phases),
one has two relations among the elements of the effective Majorana matrix. Due to the hidden relations in the four-zero texture model,
the nine parameters are not independent in contrast to the most general
case. The relations produce correlations among the phenomenological parameters. We study the correlations numerically and compare them with the relations obtained
in analytic forms.

The paper is organized as follows.
In section 2, we introduce a parametrization of effective Majorana matrix and  a CP violating observable which is a  product of Jarlskog  invariant and mass  squared differences.
In section 3, we classify $126$ four-zero textures of Dirac mass matrix into seven classes according to the configuration of zero or non-vanishing elements. Among them, we study three classes
which produce the non-zero CP violating observable and the non-vanishing lightest neutrino mass. In section 4, we outline the efficient method to study all the textures based on the classification. In section 5,  
we show the numerical results by focusing on the correlations among three CP violating phases.   In section 6, the hidden relations among the elements of the effective Majorana mass  are derived and
the correlations found in the numerical analysis are examined from a view point of the hidden relations.  
Section 7 is devoted to the summary.

\section{Model and CP violation in neutrino oscillation}
\label{sec:model}

In this section we will discuss the classification of Dirac mass matrix of the seesaw model. The effective Majorana mass matrix $m_\text{eff}$ of the seesaw model with three right-handed Majorana neutrino is written as
\begin{equation}
m_\text{eff}=-m_D\frac{1}{M}m_D^t,
\end{equation}
where $M$ is the real diagonal right-handed Majorana mass matrix and $m_D$ is the Dirac mass matrix. 
We shall use the diagonal basis in charged lepton sector. 
We name $m_{Di}  ~(i=1,2,3)$ as the magnitudes of column vectors of $m_D$ 
\begin{equation}
m_D=\mathcal{U}\left(
\begin{array}{ccc}
m_{D1} & 0 & 0 \\
0 & m_{D2} & 0 \\
0 & 0 & m_{D3}
\end{array}
\right),\quad 
\mathcal{U}=\left(
\begin{array}{ccc}
\mathcal{U}_{e1} & \mathcal{U}_{e2} & \mathcal{U}_{e3} \\
\mathcal{U}_{\mu 1} & \mathcal{U}_{\mu 2} & \mathcal{U}_{\mu 3} \\
\mathcal{U}_{\tau 1} & \mathcal{U}_{\tau 2} & \mathcal{U}_{\tau 3}
\end{array}
\right).
\end{equation}
$\mathcal{U}$ is a non-unitary matrix, which satisfies the following condition 
\begin{equation}
\sum_{\alpha=e,\mu,\tau}|\mathcal{U}_{\alpha i}|^2=1.
\label{eq:normarization}
\end{equation}
The column vectors of $\mathcal{U}$ have unit length.
Three of the 9 phases of $\mathcal{U}$ can be rotated away by flavor-basis transformation. The most general $\mathcal{U}$ involves 6 independent moduli and 6 phases.  
We define a diagonal matrix $X$ as follows
\begin{equation}
X=\left(
\begin{array}{ccc}
X_1 & 0 & 0 \\
0 & X_2 & 0 \\
0 & 0 & X_3
\end{array}
\right)
, \quad 
X_i=\frac{m_{Di}^2}{M_i}.
\label{eq:X}
\end{equation}
We rewrite the $m_\text{eff}$ in terms of $\mathcal{U}$ and $X$ \cite{Fujihara:2005pv},
\begin{equation}
m_\text{eff}=-\mathcal{U}X\mathcal{U}^t.
\label{eq:-UXUt}
\end{equation}

The CP asymmetry in neutrino oscillation, which is defined as the subtraction of transition probabilities as to neutrino and anti-neutrino, is written as follows.  
\begin{equation}
A^{CP} 
= 16 J \sin(\frac{\Delta m_{12}^2}{2E}t)\rm{sin}(\frac{\Delta m_{23}^2}{2E}t)\rm{sin}(\frac{\Delta m_{31}^2}{2E}t),
\end{equation}
where, $J$ is the Jarlskog invariant~\cite{Jarlskog:1985ht}, 
$\Delta m_{ij}^2$ $(i,j=1,2,3)$ are neutrino mass squared differences.
The Jarlskog invariant is expressed in terms of the effective mass matrix and  neutrino mass squared differences.
\begin{equation}
J
=\frac{\rm{Im}((m_{eff}m_{eff}^\dagger)_{e\mu}(m_{eff}m_{eff}^\dagger)_{\mu\tau}(m_{eff}m_{eff}^\dagger)_{\tau e})}{\Delta m_{12}^2\Delta m_{23}^2\Delta m_{31}^2}.
\label{eq:J}
\end{equation}
We define a quantity $\Delta$ as the product of Jarlskog invariant and neutrino mass squared differences, i.e. it is the numerator of Eq.(\ref{eq:J}). 
\begin{equation}
\begin{split}
\Delta
&\equiv\rm{Im}((m_{eff}m_{eff}^\dagger)_{e\mu}(m_{eff}m_{eff}^\dagger)_{\mu\tau}(m_{eff}m_{eff}^\dagger)_{\tau e})  \\
&=J\Delta m_{12}^2\Delta m_{23}^2\Delta m_{31}^2  
\label{eq:Delta}
\end{split}
\end{equation}

\section{Classification of four-zero textures of Dirac mass matrix}
\label{sec:Classification}

As we discussed above, $m_\text{eff}$ is parametrized with  15 parameters. It still has many parameters in comparison to 7 experimental observables which include 3 mixing angles, a Dirac CP violating phase, two mass squared differences, and $|(m_\text{eff})_{ee}|$. 
To reduce the number of parameters, we substitute 0s for four elements of the Dirac mass matrix $\mathcal{U}$ and assume that the other elements must not be 0.  Textures in which  $\mathcal{U}$ has five 0 elements either do not explain the CP asymmetry in neutrino oscillation or result in the minimal seesaw model with two right handed neutrinos, which causes a zero neutrino mass. The four-zeros in $\mathcal{U}$ is the most minimal texture to produce both the CP asymmetry and three non-zero mass eigenvalues of neutrino.
Taking into account of the condition that the column vectors of $\mathcal{U}$ are normalized as Eq.(\ref{eq:normarization}), any of the four-zero textures of  $\mathcal{U}$ can be expressed by four parameters after the suitable flavor-basis transformation. 
For instance, one of the four-zero textures is written as
\begin{equation}
\mathcal{U}=\left(
\begin{array}{ccc}
\cos{\theta_1}e^{i\phi_1} & \cos{\theta_2}e^{i\phi_2} & 1 \\
\sin{\theta_1} & 0 & 0 \\
0 & \sin{\theta_2} & 0
\end{array}
\right).
\label{eq.FourzeroU}
\end{equation}

The configuration of the four-zero texture in three by three matrix has ${}_9C_4=126$ different patterns. We classify them into seven classes by imposing following conditions.
\\

$(\rm{I})$ There is only one row component in which all elements are non-zero, and  $\mbox{rank}\ \mathcal{U}=3$.
\begin{equation}
\left(
\begin{array}{ccc}
* & * & * \\
* & 0 & 0 \\
0 & * & 0
\end{array}
\right)	\ ....\ 18\ \rm{patterns}
\label{class1}
\end{equation}

$(\rm{I\hspace{-1pt}I})$ There is only one column component in which all elements are non-zero, and $\mbox{rank}\ \mathcal{U}=3$.
\begin{equation}
\left(
\begin{array}{ccc}
* & * & 0 \\
* & 0 & * \\
* & 0 & 0
\end{array}
\right)	\ ....\ 18\ \rm{patterns}
\label{class2}
\end{equation}

$(\rm{I\hspace{-1pt}I\hspace{-1pt}I})$ There are one row and column components in which two elements are zero, and the common element on the both of such row and column is zero. Besides, we require that  $\mbox{rank}\ \mathcal{U}=3$. 
\begin{equation}
\left(
\begin{array}{ccc}
* & 0 & 0 \\
* & * & 0 \\
0 & * & *
\end{array}
\right)	\ ....\ 36\ \rm{patterns}
\label{class3}
\end{equation}

$(\rm{I\hspace{-1pt}V})$ There is only one column component in which all elements are zero.
\begin{equation}
\left(
\begin{array}{ccc}
* & * & 0 \\
* & * & 0 \\
* & 0 & 0
\end{array}
\right)	\ ....\ 18\ \rm{patterns}
\end{equation}

$(\rm{V})$ There is only one row component in which all elements are zero.
\begin{equation}
\left(
\begin{array}{ccc}
* & * & * \\
* & * & 0 \\
0 & 0 & 0
\end{array}
\right)	\ ....\ 18\ \rm{patterns}
\end{equation}

$(\rm{V\hspace{-1pt}I})$ There is one row component and one column component in which all elements are non-zero.
\begin{equation}
\left(
\begin{array}{ccc}
* & * & * \\
* & 0 & 0 \\
* & 0 & 0
\end{array}
\right)	\ ....\ 9\ \rm{patterns}
\end{equation}

$(\rm{V\hspace{-1pt}I\hspace{-1pt}I})$ There are one row and column components in which two elements are zero, and the common element on the both of such row and column is not zero. 
\begin{equation}
\left(
\begin{array}{ccc}
* & 0 & 0 \\
0 & * & * \\
0 & * & *
\end{array}
\right)	\ ....\ 9\ \rm{patterns}
\end{equation}				

Class $(\rm{V})$, $(\rm{V\hspace{-1pt}I})$, and $(\rm{V\hspace{-1pt}I\hspace{-1pt}I})$ lead to that $\Delta=0$. Therefore, we do not consider those three classes. Class $(\rm{I\hspace{-1pt}V})$ produces a zero mass eigenvalue of neutrino. Although the existence of one massless neutrino is still allowed at present, this class  $(\rm{I\hspace{-1pt}V})$ leads to the same effective Majorana mass matrix as that of the seesaw model with two right handed neutrinos. Taking more general assumption that all three mass eigenvalues are non-zero, we do not consider it. From the reasons above, we adopt three classes $(\rm{I})$, $(\rm{I\hspace{-1pt}I})$ and $(\rm{I\hspace{-1pt}I\hspace{-1pt}I})$.

\section{Numerical Analyses}\label{sec:num}
In this section, we perform the numerical analysis for four-zero texture models in the previous 
section.
The $m_{\rm eff}$  in models are parametrized with three mass scales  $X_i (i=1 \sim 3)$ and two angles
and two phases in   ${\cal U}$ as shown in Eq.(\ref{eq.FourzeroU}).
To determine $X_i (i=1 \sim 3)$, we consider the eigenvalue equation for neutrino mass squared \cite{Morozumi:2005aa},
\begin{eqnarray}
\det(m_{\rm eff} m^\dagger_{\rm eff}-\lambda I)=0.
\label{eq:eigenvalue}
\end{eqnarray}
$I$ is the three by three unite matrix and $\lambda$'s correspond to mass eigenvalues. 
The equation leads to three set of constraints on $X_i$,
\begin{eqnarray}
&&\tr(m_{\rm eff} m^\dagger_{\rm eff})=\sum_{i=1}^3 m_i^2,
\label{eq1:constraintX}
\\
&&\det(m_{\rm eff} m^\dagger_{\rm eff})=m_1^2m_2^2 m_3^2 ,
\label{eq2:constraintX}
\\
&&\frac{\{\tr(m_{\rm eff} m^\dagger_{\rm eff})\}^2-\tr\{(m_{\rm eff} m^\dagger_{\rm eff})^2\}}{2}=m_1^2 m_2^2+m_2^2m_3^2+m_3^2m_1^2,
%
\label{eq3:constraintX}
\end{eqnarray}
where $m_i (i=1 \sim 3)$ are mass eigenvalues of neutrinos.
The left-hand side of Eqs.\eqref{eq1:constraintX} -\eqref{eq3:constraintX} are written in terms of 
model parameters: $X_i$
, angles $\theta_1, \theta_2$ and phases $\phi_1, \phi_2$. The numerical values of the right-hand side
are fixed with the neutrino mass squared differences for a given value of  the lightest neutrino mass
and hierarchy (normal or inverted ) of neutrino mass spectrum. 
We randomly generate a set of values of $\Delta m_{\rm sol.}^2$ and $\Delta m_{\rm atm.}^2$ 
within $3$ standard deviations from the mean of the experimental values~\cite{Gonzalez-Garcia:2014bfa}. 
Assigning the numerical values for angles and phases randomly from $-\pi$ to $\pi$,
one can determine the sets of $X_i$ which satisfy the constraints  Eqs.\eqref{eq1:constraintX} -\eqref{eq3:constraintX}.
$m_{\rm eff}$ is reconstructed in terms of the numerically allocated parameters  $\theta_1, \theta_2,\phi_1, \phi_2 $, the lightest neutrino
mass $m_1$($m_3$) for normal (inverted) hierarchy, and solved $X$s. 
By using the obtained $m_{\rm eff}$, we compute mixing angles $\theta_{12},\theta_{23}, \theta_{31}$ and three CP phases $\delta, \alpha_{21},
\alpha_{31}$. If all three mixing angles are within $3$ standard deviations from the mean of the experimental values, we take the parameters as a possible model.
CP phases and the correlations among them can be predictions.  We collect the set of the parameters
which produce the appropriate mixing angles by repeating this procedure.
In principle, the procedure explained above can be applied to all the textures in class (I), $(\rm{I\hspace{-1pt}I})$ and $(\rm{I\hspace{-1pt}I\hspace{-1pt}I})$. 
However, we do not have to repeat the procedure for all textures in a class individually. By using some relations of 
parameters of two different textures in a class, we can save labor to examine them.

The configurations of all textures in a specific class are related to each other by exchanging their column and rows.
Let us define the matrices $Q$ and $P$ which correspond to replacements of row and column components, respectively.    
\begin{equation}
P,Q = \left(
\begin{array}{ccc}
1 & 0 & 0 \\
0 & 1 & 0 \\
0 & 0 & 1
\end{array}
\right),
\left(
\begin{array}{ccc}
1 & 0 & 0 \\
0 & 0 & 1 \\
0 & 1 & 0
\end{array}
\right),
\left(
\begin{array}{ccc}
0 & 1 & 0 \\
1 & 0 & 0 \\
0 & 0 & 1
\end{array}
\right),
\left(
\begin{array}{ccc}
0 & 1 & 0 \\
0 & 0 & 1 \\
1 & 0 & 0
\end{array}
\right),
\left(
\begin{array}{ccc}
0 & 0 & 1 \\
1 & 0 & 0 \\
0 & 1 & 0
\end{array}
\right),
\left(
\begin{array}{ccc}
0 & 0 & 1 \\
0 & 1 & 0 \\
1 & 0 & 0
\end{array}
\right).
\end{equation}	
If we only consider the position of zero and non-zero elements on the Dirac mass matrix, the configuration between a texture  $\mathcal{U}$ and another texture $\mathcal{U}'$ in the same class is expressed as 

\begin{equation}
\mathcal{U}'=Q\mathcal{U}P.
\label{eq:QUP}
\end{equation}	
Taking the suitable $P$ and $Q$, any two textures in the same class are related to each other by Eq.(\ref{eq:QUP}).
Let us define $X'$ as
\begin{equation}
X'=P^tXP.
\label{eq:PtXP}
\end{equation}	
The $m'_{\rm{eff}}$, which is constructed by $\mathcal{U}'$ and $X'$ through the definition Eq.(\ref{eq:-UXUt}) in stead of  $\mathcal{U}$ and $X$, is expressed in terms of $m_{\rm{eff}}$ and $Q$.
\begin{equation}
\begin{split}
m'_{\rm{eff}}&=-\mathcal{U}'X'(\mathcal{U}')^t \\
&=Qm_{\rm{eff}}Q^T.  
\label{eq:QmeffQt}
\end{split}
\end{equation}	
The $m'_{\rm{eff}}$ satisfies the same constraints Eqs.\eqref{eq1:constraintX}-\eqref{eq3:constraintX}.  
We introduce an unitary matrix $V$ which diagonalizes $m_{\rm{eff}}m_{\rm{eff}}^\dagger$,
\begin{equation}
V^\dagger(m_{\rm{eff}}m_{\rm{eff}}^\dagger)V=
\left(
\begin{array}{ccc}
m_1^2 & 0 & 0 \\
0 & m_2^2 & 0 \\
0 & 0 & m_3^2
\end{array}
\right).
\end{equation}
While the mixing matrix $V'$, which diagonalizes $m'_{\rm{eff}}(m'_{\rm{eff}})^\dagger$, is 
\begin{equation}
V'=QV.
\label{eq:QV}
\end{equation}
If we once do the analysis as for a texture, we obtain another $m'_{\rm{eff}}$ by arranging the elements of $m_{\rm{eff}}$. The other effective mass matrices in the same class are calculated out without solving extra eigenvalue equations. 
Since it reduces the processing loads of solving eigenvalue equations, the method is much more efficient for numerical analysis.

Next, we investigate the classification of Eq.(\ref{class1})-(\ref{class2}) in detail.
In each class  of  (I)-(III), we further classify the textures into subclasses.
The 18 textures of the class $(\rm{I})$ are classified into three subclasses called class $(\rm{I})$-A, $(\rm{I})$-B and $(\rm{I})$-C. The Dirac matrices which belong to each subclass  have a row with all non-zero elements.     
In matrices of class $(\rm{I})$-A, all elements of the first row
do not vanish and in class $(\rm{I})$-B, all of the elements of the second row does not vanish. In class $(\rm{I})$-C, all of them in the third row  have non-zero values. These subclasses can be related to each other by applying the permutation matrix $Q$
on $\mathcal{U}$. 
By multiplying the permutation matrix $P$ on 
a Dirac matrix  $\cal U$ in each subclass,
\begin{equation}
\mathcal{U}''=\mathcal{U}P,
\label{eq:UP}
\end{equation}
one can generate all six different matrices which belong to the same subclass. 
Eq.(\ref{eq:UP}) corresponds to the specific case of Eq.(\ref{eq:QUP}) such that $Q$ is identical to the unit matrix, and indicates the replacements of columns.
In this way,  we obtain 6 matrices in the 
 class $(\rm{I})$-A with the following textures, 
\begin{equation}
\left(
\begin{array}{ccc}
* & * & * \\
* & 0 & 0 \\
0 & * & 0
\end{array}
\right),
\left(
\begin{array}{ccc}
* & * & * \\
* & 0 & 0 \\
0 & 0 & *
\end{array}
\right),
\left(
\begin{array}{ccc}
* & * & * \\
0 & * & 0 \\
* & 0 & 0
\end{array}
\right),
\left(
\begin{array}{ccc}
* & * & * \\
0 & * & 0 \\
0 & 0 & *
\end{array}
\right),
\left(
\begin{array}{ccc}
* & * & * \\
0 & 0 & * \\
* & 0 & 0
\end{array}
\right),
\left(
\begin{array}{ccc}
* & * & * \\
0 & 0 & * \\
0 & * & 0
\end{array}
\right).
\label{eq:I-A}
\end{equation}
The effective matrix constructed from $(\mathcal{U}'',X^\prime)$ is the same as that from $({\mathcal U},X)$ and it is diagonalized by the same matrix $V$.
If one finds real solutions of Eq.(\ref{eq:eigenvalue}) for $X$s and if obtained mixing angles are consistent with the experimental results from neutrino oscillation, it is confirmed that the other 5 textures also have model parameters satisfying with the constraints from mass eigenvalues and mixing angles.
On the other hand, if one texture do not give any correct mixing angles, there is no hope that the others would realize the experimental results either.

Let us define subclasses of Dirac matrices called as $(\rm{I})$-B and $(\rm{I})$-C. The matrices in $(\rm{I})$-B are obtained by multiplying   
those in subclass $(\rm{I})$-A by
\begin{equation}
Q=\left(
\begin{array}{ccc}
0 & 0 & 1 \\
1 & 0 & 0 \\
0 & 1 & 0
\end{array}
\right).
\end{equation} In the same way, the matrices in $(\rm{I})$-C are obtained by multiplying those in $(\rm{I})$-A by matrix
\begin{equation}
Q=\left(
\begin{array}{ccc}
0 & 1 & 0 \\
0 & 0 & 1 \\
1 & 0 & 0
\end{array}
\right).
\end{equation}

Any subclass has 6 textures related by Eq.(\ref{eq:UP}).
We carry out the same manner of classification for class $(\rm{I\hspace{-1pt}I})$ as we have done for class $(\rm{I})$.
Class $(\rm{I\hspace{-1pt}I\hspace{-1pt}I})$ contains 36 textures, therefore it is categorized into 6 subclasses from $(\rm{I\hspace{-1pt}I\hspace{-1pt}I})$-A to $(\rm{I\hspace{-1pt}I\hspace{-1pt}I})$-F.
We arrange the subclasses in Table \ref{table:subclasses}, by picking up one representation for each subclass.

\begin{table}[htb]
	\begin{center}
		\begin{tabular}{|l|c|c|c|} \hline
			& class $(\rm{I})$ & class $(\rm{I\hspace{-1pt}I})$ & class $(\rm{I\hspace{-1pt}I\hspace{-1pt}I})$ \\ \hline 
			A & $\left(
			\begin{array}{ccc}
			* & * & * \\
			* & 0 & 0 \\
			0 & * & 0
			\end{array}
			\right)$ 
			& $
			\left(
			\begin{array}{ccc}
			* & * & 0 \\
			* & 0 & * \\
			* & 0 & 0
			\end{array}
			\right)$ 
			& $\left(
			\begin{array}{ccc}
			* & 0 & 0 \\
			* & * & 0 \\
			0 & * & *
			\end{array}
			\right)$ \\
			B & $\left(
			\begin{array}{ccc}
			0 & * & 0 \\
			* & * & * \\
			* & 0 & 0 
			\end{array}
			\right)$ &  $\left(
			\begin{array}{ccc}
			* & 0 & 0 \\
			* & * & 0 \\
			* & 0 & * 
			\end{array}
			\right)$  & $\left(
			\begin{array}{ccc}
			0 & * & * \\
			* & 0 & 0 \\
			* & * & 0 
			\end{array}
			\right)$ \\
			C &$\left(
			\begin{array}{ccc}
			* & 0 & 0 \\
			0 & * & 0 \\
			* & * & * 
			\end{array}
			\right)$  & $\left(
			\begin{array}{ccc}
			* & 0 & * \\
			* & 0 & 0 \\
			* & * & 0 
			\end{array}
			\right)$ & $\left(
			\begin{array}{ccc}
			* & * & 0 \\
			0 & * & * \\
			* & 0 & 0 
			\end{array}
			\right)$ \\ 
			D &  &  &  $\left(
			\begin{array}{ccc}
			* & * & 0 \\
			* & 0 & 0 \\
			0 & * & *
			\end{array}
			\right)$ \\
			E &  &  &  $\left(
			\begin{array}{ccc}
			* & 0 & 0 \\
			0 & * & * \\
			* & * & 0
			\end{array}
			\right)$ \\
			F &  &  &  $\left(
			\begin{array}{ccc}
			0 & * & * \\
			* & * & 0 \\
			* & 0 & 0
			\end{array}
			\right)$ \\ \hline
		\end{tabular}
			\caption{The classification of textures into some subclasses}
			\label{table:subclasses}
	\end{center}
\end{table}

\begin{table}[htb]
	\begin{center}
		\begin{tabular}{|l|c|c|c|} \hline
			& class $(\rm{I})$ & class $(\rm{I\hspace{-1pt}I})$ & class $(\rm{I\hspace{-1pt}I\hspace{-1pt}I})$ \\ \hline 
			A & $\left(
			\begin{array}{ccc}
			\sin{\theta_1}e^{i\phi_1} & \sin{\theta_2}e^{i\phi_2} & 1 \\
			\cos{\theta_1} & 0 & 0 \\
			0 & \cos{\theta_2} & 0
			\end{array}
			\right)$ 
			& $
			\left(
			\begin{array}{ccc}
			\sin{\theta_1}\cos{\theta_2} & e^{i\phi_1} & 0 \\
			\sin{\theta_1}\sin{\theta_2} & 0 & e^{i\phi_2} \\
			\cos{\theta_1} & 0 & 0
			\end{array}
			\right)$ 
			& $\left(
			\begin{array}{ccc}
			\sin{\theta_1} & 0 & 0 \\
			\cos{\theta_1} & \sin{\theta_2}e^{i\phi_1} & 0 \\
			0 & \cos{\theta_2}e^{i\phi_2} & 1
			\end{array}
			\right)$ \\
		    &	&Fig.\ref{fig.NH2ACPa21}, Fig.\ref{fig.NH2ACPa31}(NH)&
		    	Fig.\ref{fig.NH3ACPa21},Fig.\ref{fig.NH3ACPa31}(NH) \\
& & &
		    	Fig.\ref{fig.IH3ACPa21},
		     Fig.\ref{fig.IH3ACPa31}(IH)
		    \\
			\hline 
			B & $\left(
			\begin{array}{ccc}
			0 & \cos{\theta_2} & 0 \\
			\sin{\theta_1}e^{i\phi_1} & \sin{\theta_2}e^{i\phi_2} & 1 \\
			\cos{\theta_1} & 0 & 0 
			\end{array}
			\right)$ &  $\left(
			\begin{array}{ccc}
			\cos{\theta_1} & 0 & 0 \\
			\sin{\theta_1}\cos{\theta_2} & e^{i\phi_1} & 0 \\
			\sin{\theta_1}\sin{\theta_2} & 0 & e^{i\phi_2} 
			\end{array}
			\right)$  & $\left(
			\begin{array}{ccc}
			0 & \cos{\theta_2}e^{i\phi_2} & 1  \\
			\sin{\theta_1} & 0 & 0 \\
			\cos{\theta_1} & \sin{\theta_2}e^{i\phi_1} & 0 
			\end{array}
			\right)$ \\
			&Fig.\ref{fig.IH1BCPa21}, Fig.\ref{fig.IH1BCPa31}(IH),& &Fig.\ref{fig.IH3BCPa21}, Fig.\ref{fig.IH3BCPa31}(IH)\\
			\hline 
			C &$\left(
			\begin{array}{ccc}
			\cos{\theta_1} & 0 & 0 \\
			0 & \cos{\theta_2} & 0 \\
			\sin{\theta_1}e^{i\phi_1} & \sin{\theta_2}e^{i\phi_2} & 1 
			\end{array}
			\right)$  & $\left(
			\begin{array}{ccc}
			\sin{\theta_1}\sin{\theta_2} & 0 & e^{i\phi_2} \\
			\cos{\theta_1} & 0 & 0 \\
			\sin{\theta_1}\cos{\theta_2} & e^{i\phi_1} & 0 
			\end{array}
			\right)$ & $\left(
			\begin{array}{ccc}
			\cos{\theta_1} & \sin{\theta_2}e^{i\phi_1} & 0 \\
			0 & \cos{\theta_2}e^{i\phi_2} & 1  \\
			\sin{\theta_1} & 0 & 0 
			\end{array}
			\right)$ \\ 
		    &Fig.\ref{fig.IH1CCPa21}, Fig.\ref{fig.IH1CCPa31}(IH)	&
		    Fig.\ref{fig.NH2CCPa21}, Fig.\ref{fig.NH2CCPa31}(NH) &
		    \\
			\hline 
			D &  &  &  $\left(
			\begin{array}{ccc}
			\cos{\theta_1} & \sin{\theta_2}e^{i\phi_1} & 0 \\
			\sin{\theta_1} & 0 & 0 \\
			0 & \cos{\theta_2}e^{i\phi_2} & 1
			\end{array}
			\right)$ \\
			& & & \\
			\hline 
			E &  &  &  $\left(
			\begin{array}{ccc}
			\sin{\theta_1} & 0 & 0 \\
			0 & \cos{\theta_2}e^{i\phi_2} & 1  \\
			\cos{\theta_1} & \sin{\theta_2}e^{i\phi_1} & 0 
			\end{array}
			\right)$ \\
			& & & Fig.\ref{fig.IH3ECPa21}, Fig.\ref{fig.IH3ECPa31}(IH)\\
			\hline 
			F &  &  &  $\left(
			\begin{array}{ccc}
			0 & \cos{\theta_2}e^{i\phi_2} & 1 \\
			\cos{\theta_1} & \sin{\theta_2}e^{i\phi_1} & 0 \\
			\sin{\theta_1} & 0 & 0
			\end{array}
			\right)$ \\
			& & & Fig.\ref{fig.IH3FCPa21}, Fig.\ref{fig.IH3FCPa31}(IH)
			\\ \hline
		\end{tabular}
		\caption{The textures of subclasses with concrete model parameters ($\theta_1$,$\theta_2$,$\phi_1$, and $\phi_2$) and the corresponding figure numbers}
		\label{table:subclassesparameters}
	\end{center}
\end{table}

\section{results}
\label{sec:results}
\begin{figure}
\begin{center}
	\includegraphics[width=7cm]{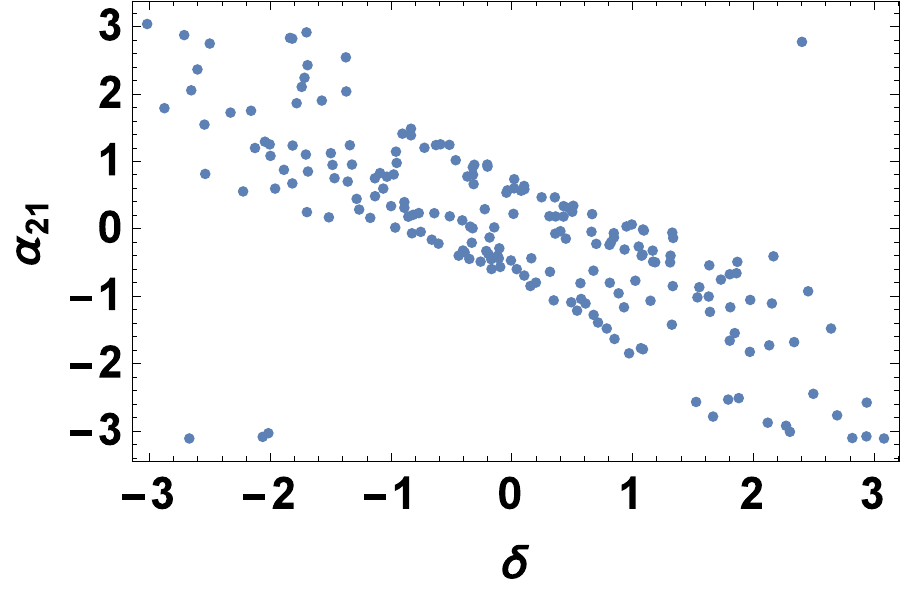}
	\caption{$\delta$ vs $\alpha_{21}$ $(\rm{I\hspace{-1pt}I\hspace{-1pt}I})$-A in NH}
	\label{fig.NH3ACPa21}
\end{center}
\end{figure}

\begin{figure}[h!]
	\begin{tabular}{ccc}
		\begin{minipage}{0.475\hsize}
			\includegraphics[width=\linewidth]{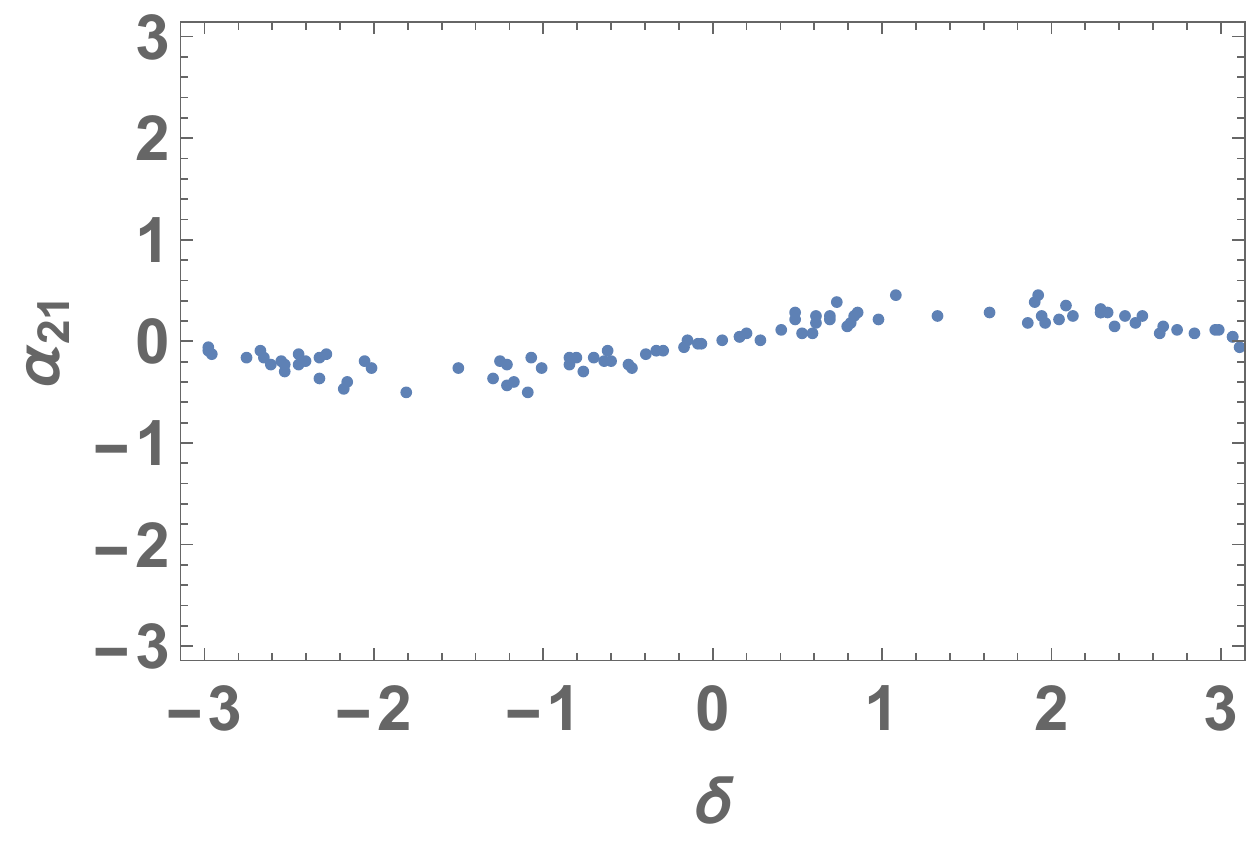}
			\caption{$\delta$ vs $\alpha_{21}$ $(\rm{I\hspace{-1pt}I})$-A in NH}
			\label{fig.NH2ACPa21}
		\end{minipage}
		\phantom{=}
		\begin{minipage}{0.475\linewidth}
			\vspace{0cm}
			\includegraphics[width=0.975\linewidth]{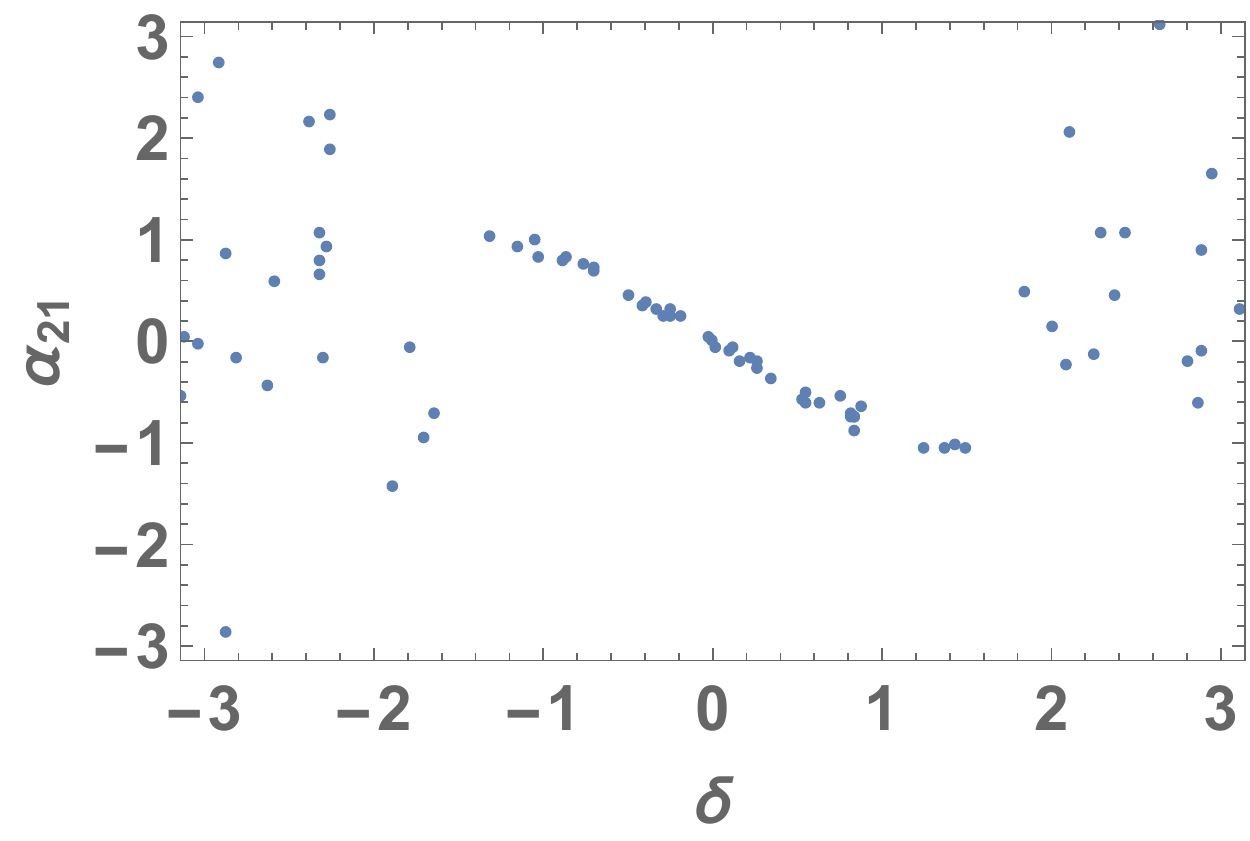}
			\caption{$\delta$ vs $\alpha_{21}$ $(\rm{I\hspace{-1pt}I})$-C in NH}
			\label{fig.NH2CCPa21}
		\end{minipage}
	\end{tabular}
\end{figure}

\begin{figure}[h!]
	\begin{tabular}{ccc}
		\begin{minipage}{0.475\hsize}
			\includegraphics[width=\linewidth]{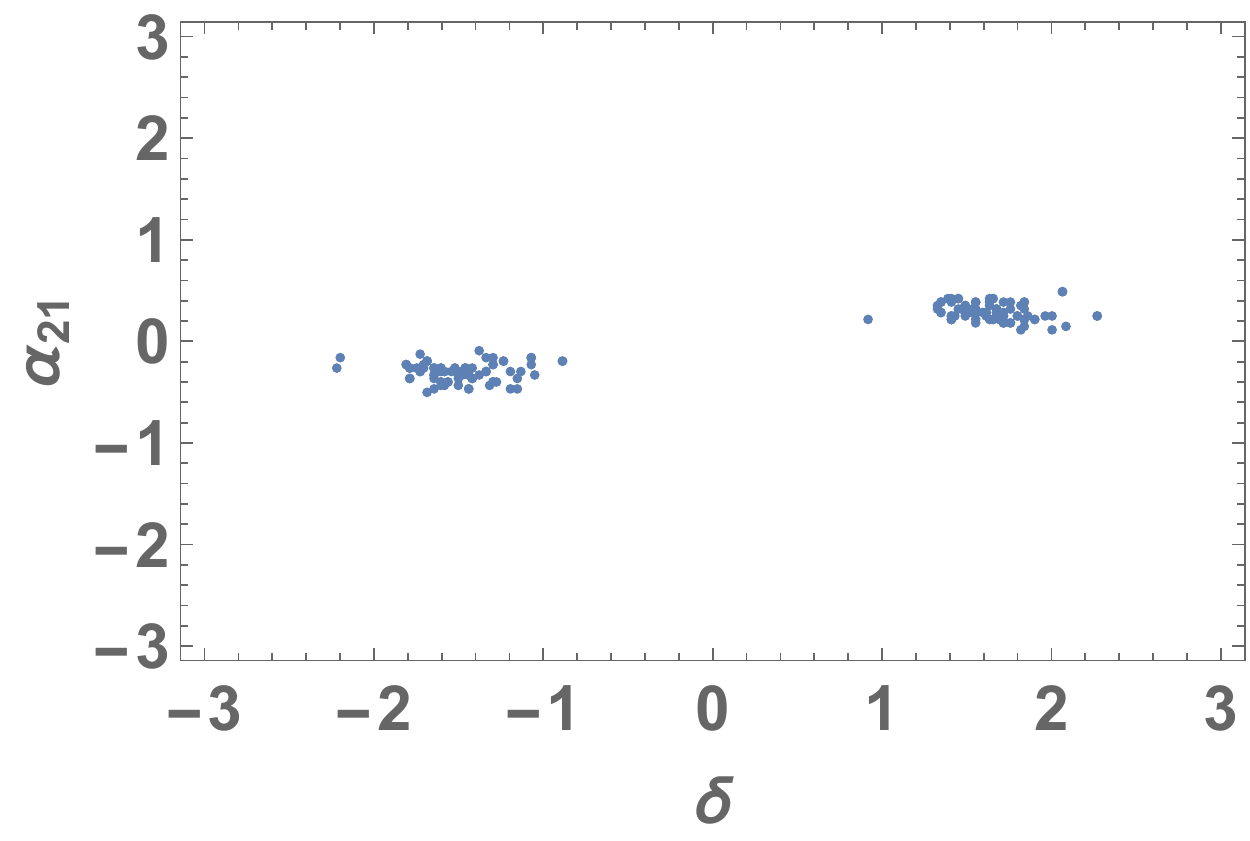}
			\caption{$\delta$ vs $\alpha_{21}$ $(\rm{I})$-B in IH}
			\label{fig.IH1BCPa21}
		\end{minipage}
		\phantom{=}
		\begin{minipage}{0.475\linewidth}
			\vspace{0cm}
			\includegraphics[width=0.975\linewidth]{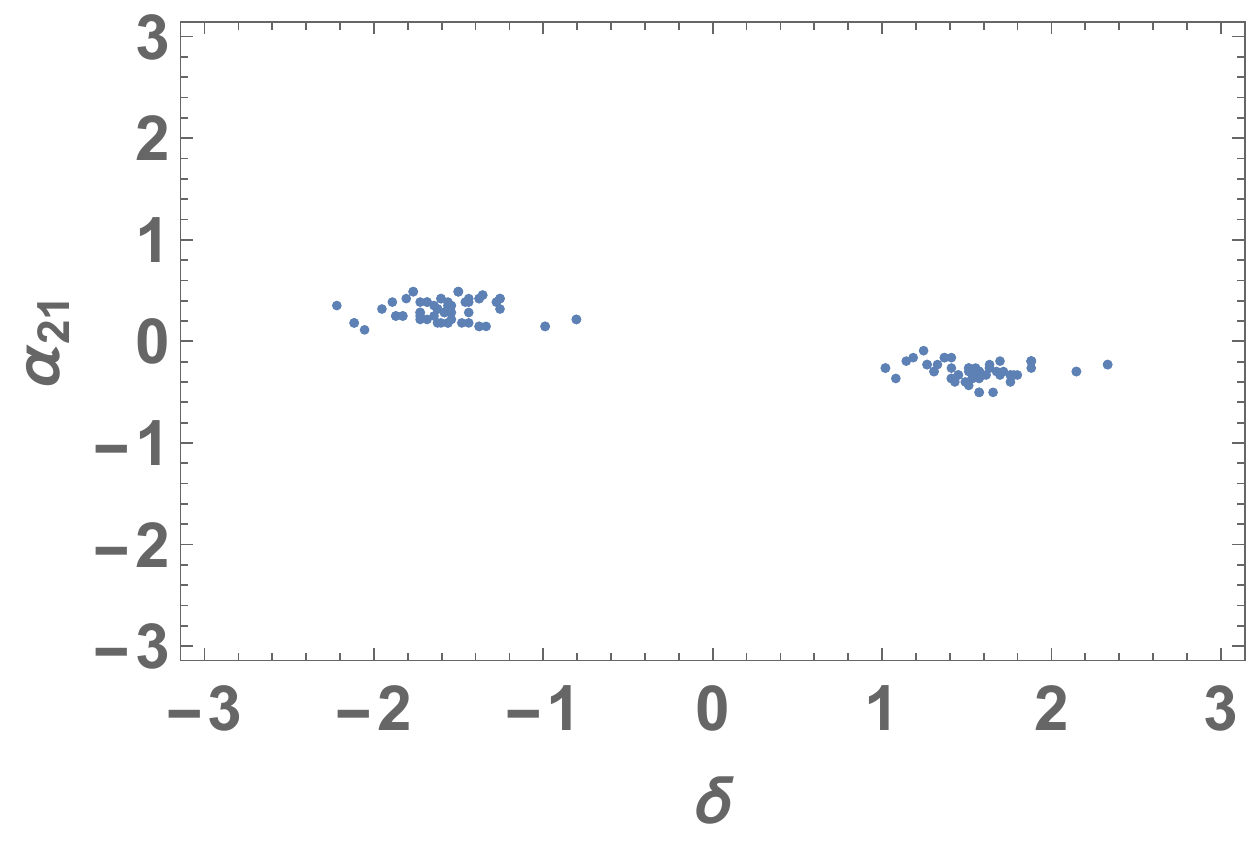}
			\caption{$\delta$ vs $\alpha_{21}$ $(\rm{I})$-C in IH}
			\label{fig.IH1CCPa21}
		\end{minipage}
	\end{tabular}
\end{figure}

\begin{figure}[h!]
	\begin{tabular}{ccc}
		\begin{minipage}{0.475\hsize}
			\includegraphics[width=\linewidth]{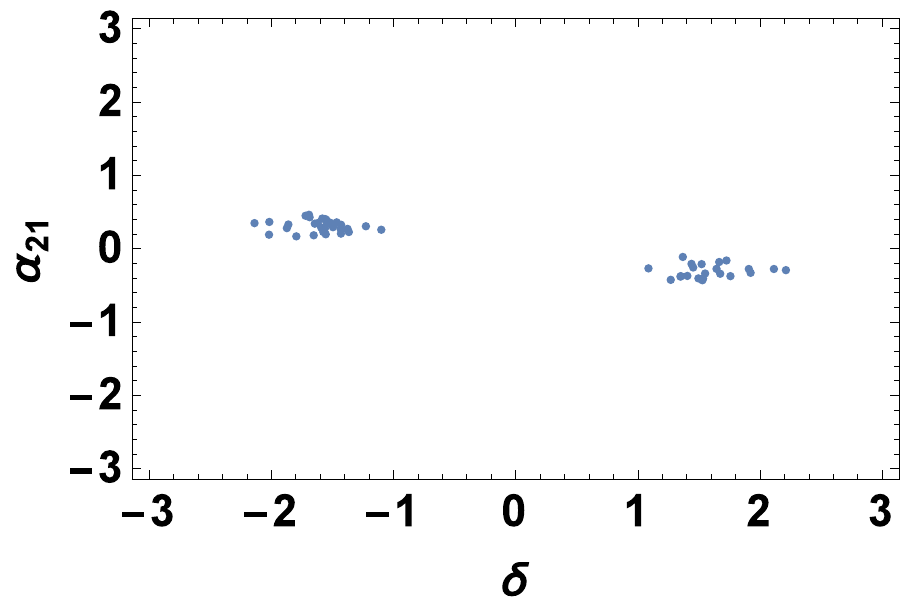}
			\caption{$\delta$ vs $\alpha_{21}$ $(\rm{I\hspace{-1pt}I\hspace{-1pt}I})$-A in IH}
			\label{fig.IH3ACPa21}
		\end{minipage}
		\phantom{=}
		\begin{minipage}{0.475\linewidth}
			\vspace{0cm}
			\includegraphics[width=0.975\linewidth]{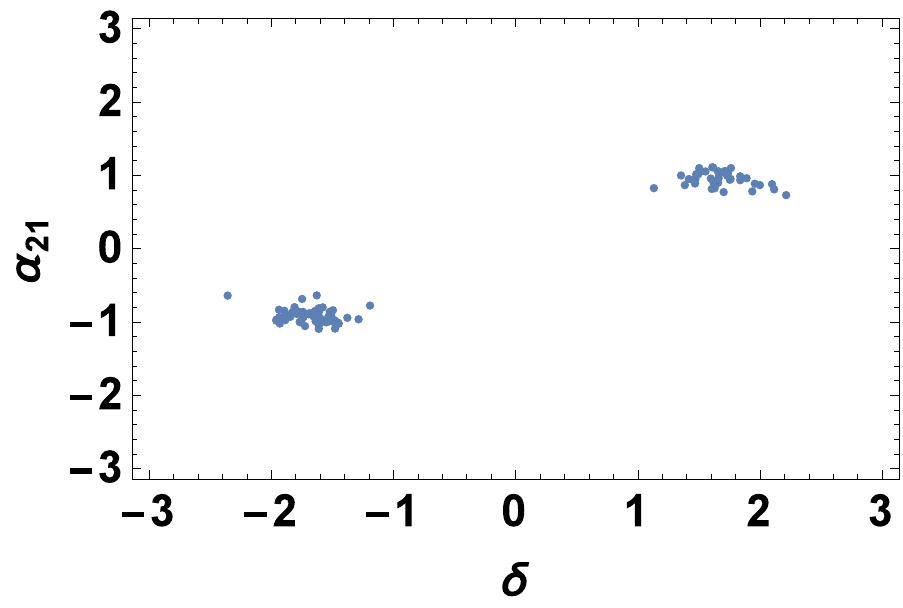}
			\caption{$\delta$ vs $\alpha_{21}$ $(\rm{I\hspace{-1pt}I\hspace{-1pt}I})$-B in IH}
			\label{fig.IH3BCPa21}
		\end{minipage}
	\end{tabular}
\end{figure}

\begin{figure}[h!]
	\begin{tabular}{ccc}
		\begin{minipage}{0.475\hsize}
			\includegraphics[width=\linewidth]{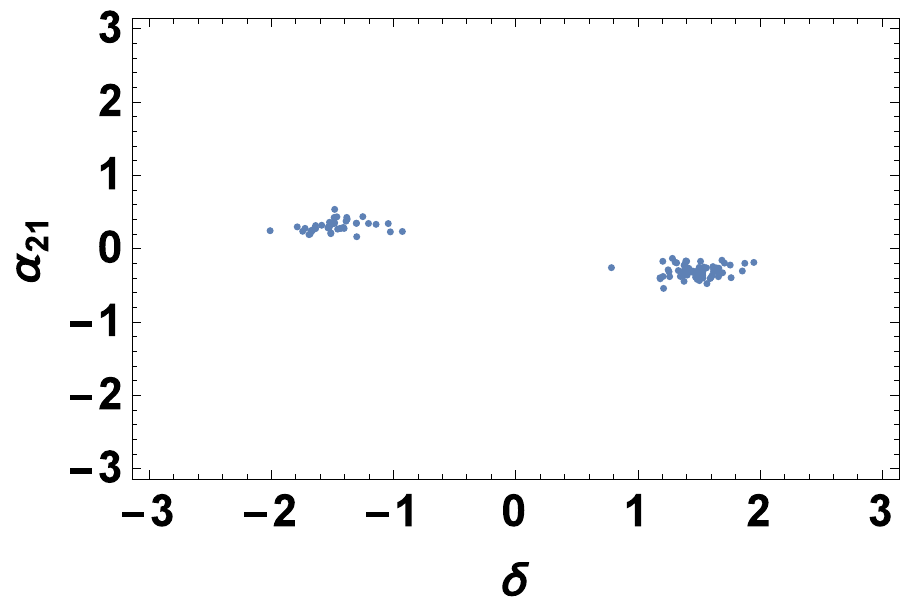}
			\caption{$\delta$ vs $\alpha_{21}$ $(\rm{I\hspace{-1pt}I\hspace{-1pt}I})$-E in IH}
			\label{fig.IH3ECPa21}
		\end{minipage}
		\phantom{=}
		\begin{minipage}{0.475\linewidth}
			\vspace{0cm}
			\includegraphics[width=0.975\linewidth]{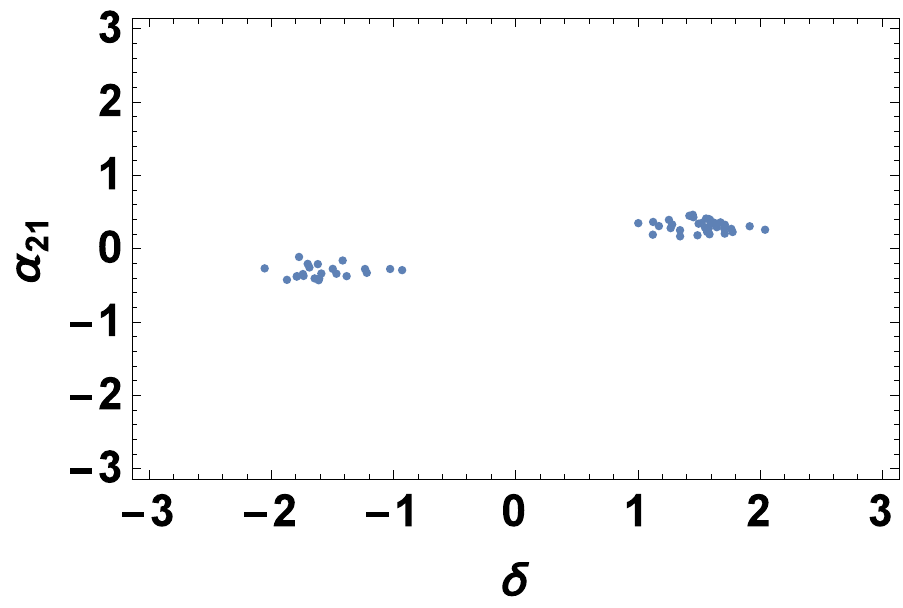}
			\caption{$\delta$ vs $\alpha_{21}$ $(\rm{I\hspace{-1pt}I\hspace{-1pt}I})$-F in IH}
			\label{fig.IH3FCPa21}
		\end{minipage}
	\end{tabular}
\end{figure}

We study the correlations among CP violating phases $\delta$, $\alpha_{21}$, and $\alpha_{31}$ as for all textures which we have classified on the Table (\ref{table:subclasses}).
In Figs.(\ref{fig.NH3ACPa21}-\ref{fig.IH3FCPa21}) we show some results that have a characteristic correlation as below. 
The cases with few data are not shown. 
We do not show the cases such that there is no strong correlation.

We first show them for normal hierarchical case.
For subclass $(\rm{I\hspace{-1pt}I\hspace{-1pt}I})$-A, whose correlation is shown in Fig.(\ref{fig.NH3ACPa21}), the dots of $(\delta,\alpha_{21})$ are distributed in a belt on a diagonal line from one corner to the opposite side. 
Some dots are also found near the other corners.
Fig.(\ref{fig.NH2ACPa21}) shows the correlation for the subclass $(\rm{I\hspace{-1pt}I})$-A. $\delta$ takes any values form $-\pi$ to $\pi$, while $\alpha_{21}$ appears near the value of 0 independent of $\delta$.
In subclass $(\rm{I\hspace{-1pt}I})$-C in Fig.(\ref{fig.NH2CCPa21}), $\alpha_{21}$ is roughly proportional to $-\delta$ in a range where both $\alpha_{21}$ and $\delta$ lie from -1 to 1. Strong correlation is not found outside the range. 


\begin{figure}
	\begin{center}
		\includegraphics[width=7cm]{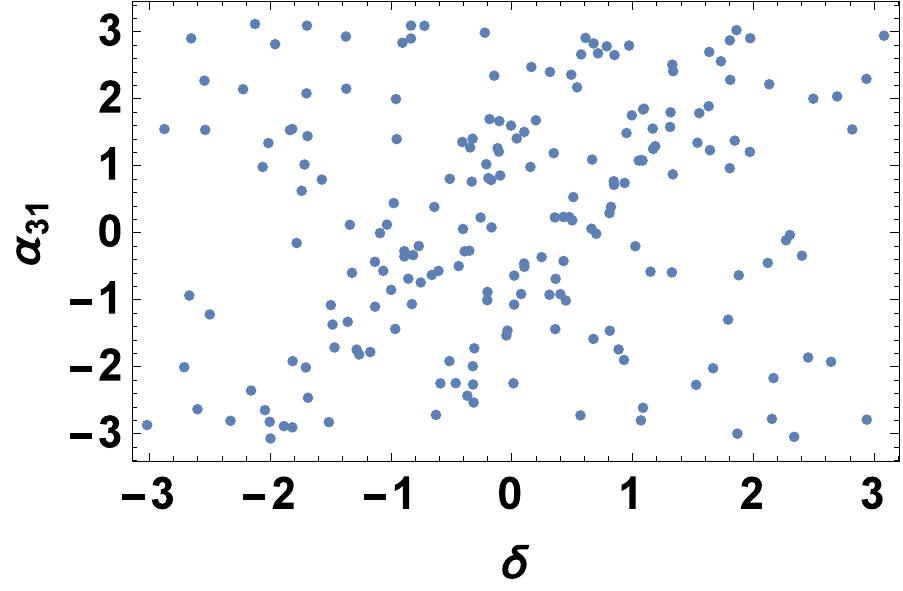}
		\caption{$\delta$ vs $\alpha_{31}$ $(\rm{I\hspace{-1pt}I\hspace{-1pt}I})$-A in NH}
		\label{fig.NH3ACPa31}
	\end{center}
\end{figure}
\begin{figure}[h!]
	\begin{tabular}{ccc}
		\begin{minipage}{0.475\hsize}
			\includegraphics[width=\linewidth]{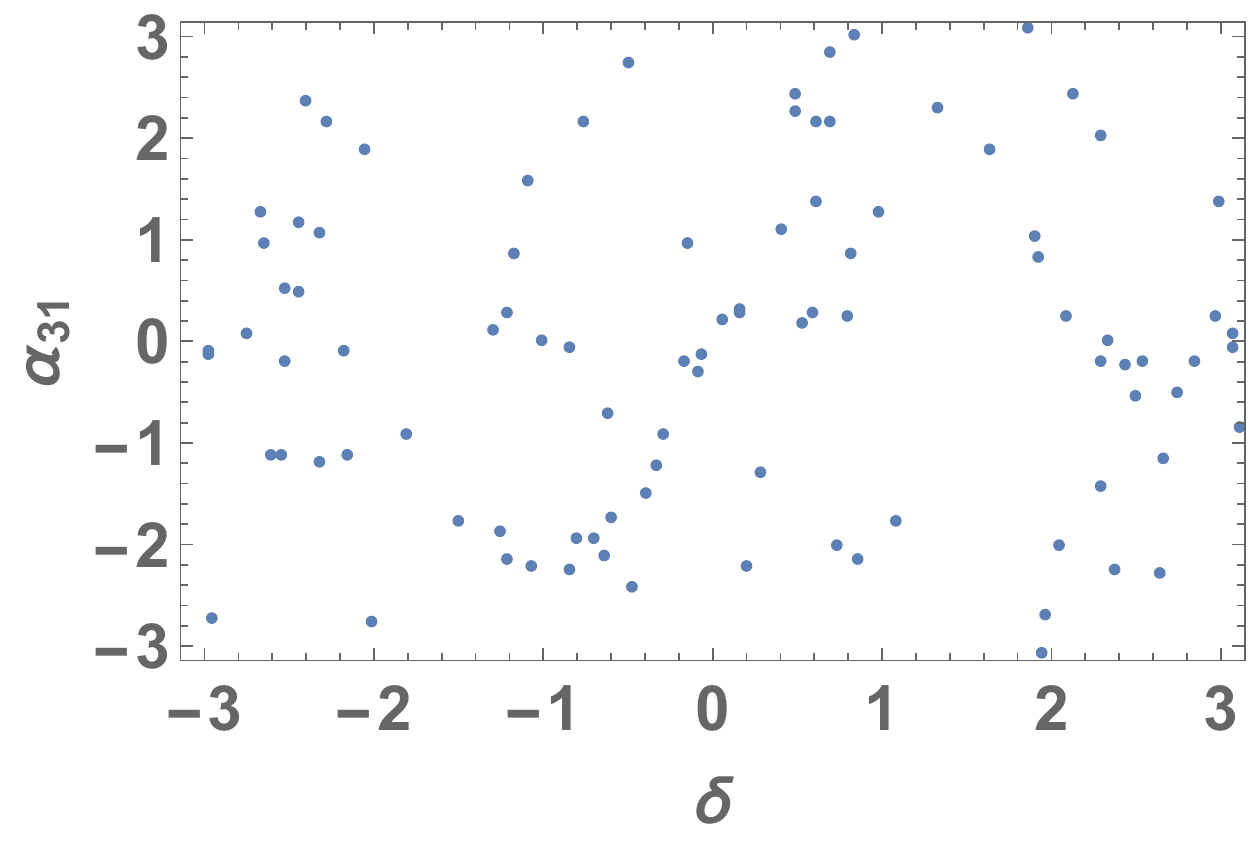}
			\caption{$\delta$ vs $\alpha_{31}$ $(\rm{I\hspace{-1pt}I})$-A in NH}
			\label{fig.NH2ACPa31}
		\end{minipage}
		\phantom{=}
		\begin{minipage}{0.475\linewidth}
			\vspace{0cm}
			\includegraphics[width=0.975\linewidth]{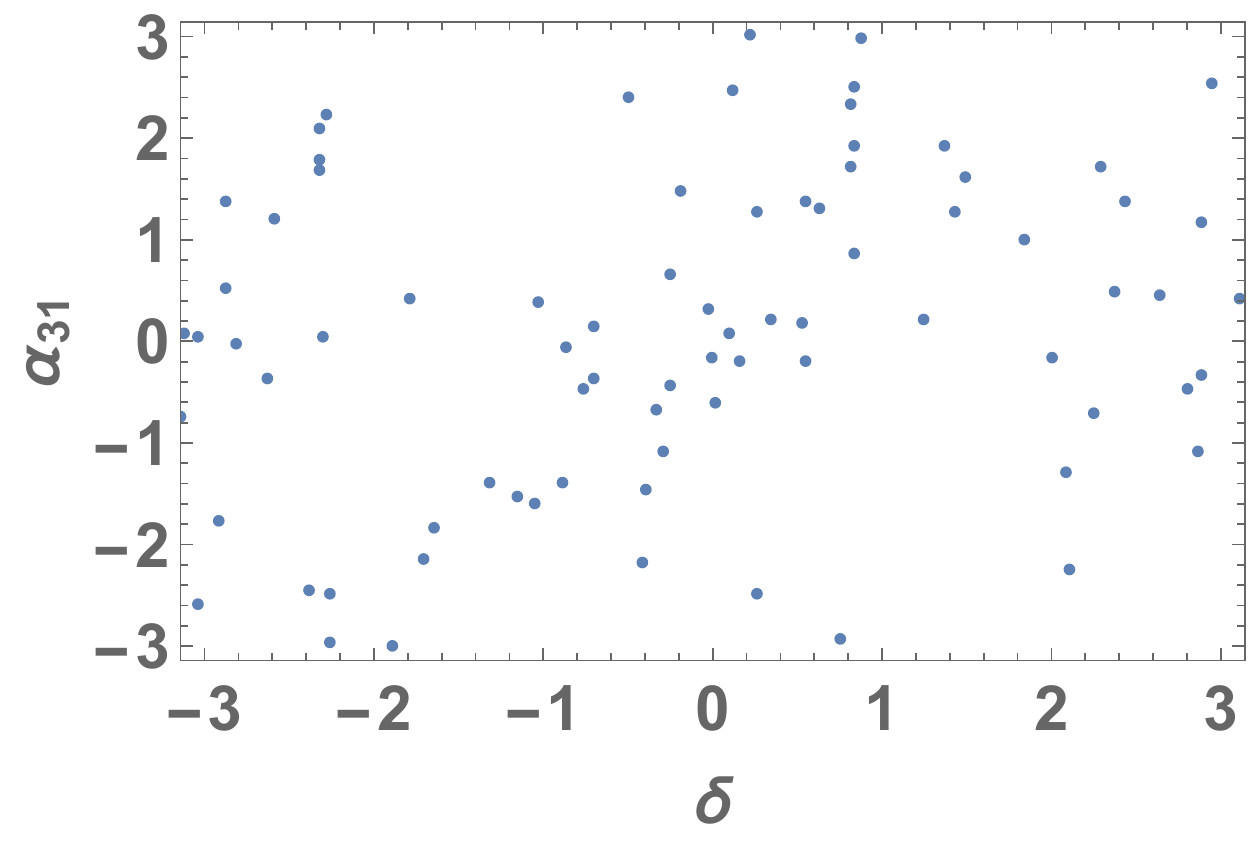}
			\caption{$\delta$ vs $\alpha_{31}$ $(\rm{I\hspace{-1pt}I})$-C in NH}
			\label{fig.NH2CCPa31}
		\end{minipage}
	\end{tabular}
\end{figure}

\begin{figure}[h!]
	\begin{tabular}{ccc}
		\begin{minipage}{0.475\hsize}
			\includegraphics[width=\linewidth]{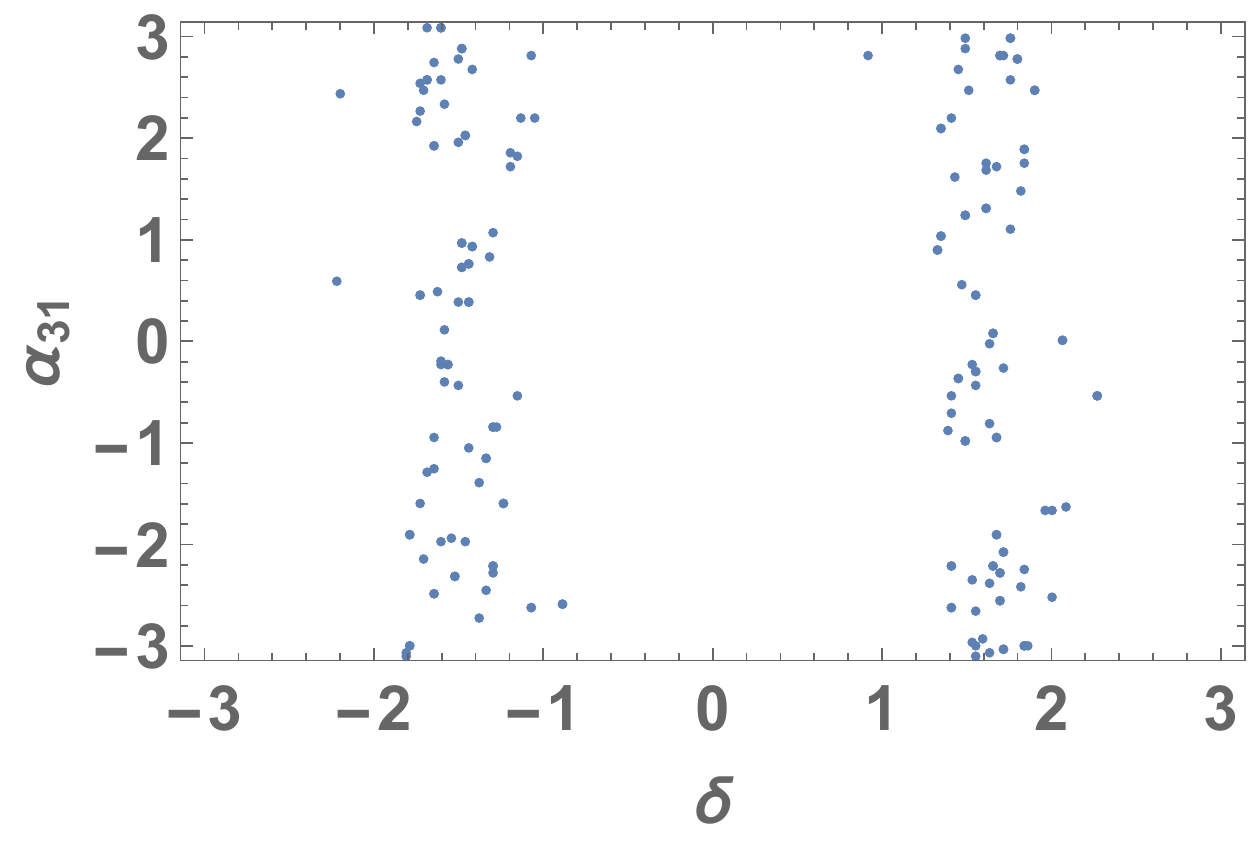}
			\caption{$\delta$ vs $\alpha_{31}$ $(\rm{I})$-B in IH}
			\label{fig.IH1BCPa31}
		\end{minipage}
		\phantom{=}
		\begin{minipage}{0.475\linewidth}
			\vspace{0cm}
			\includegraphics[width=0.975\linewidth]{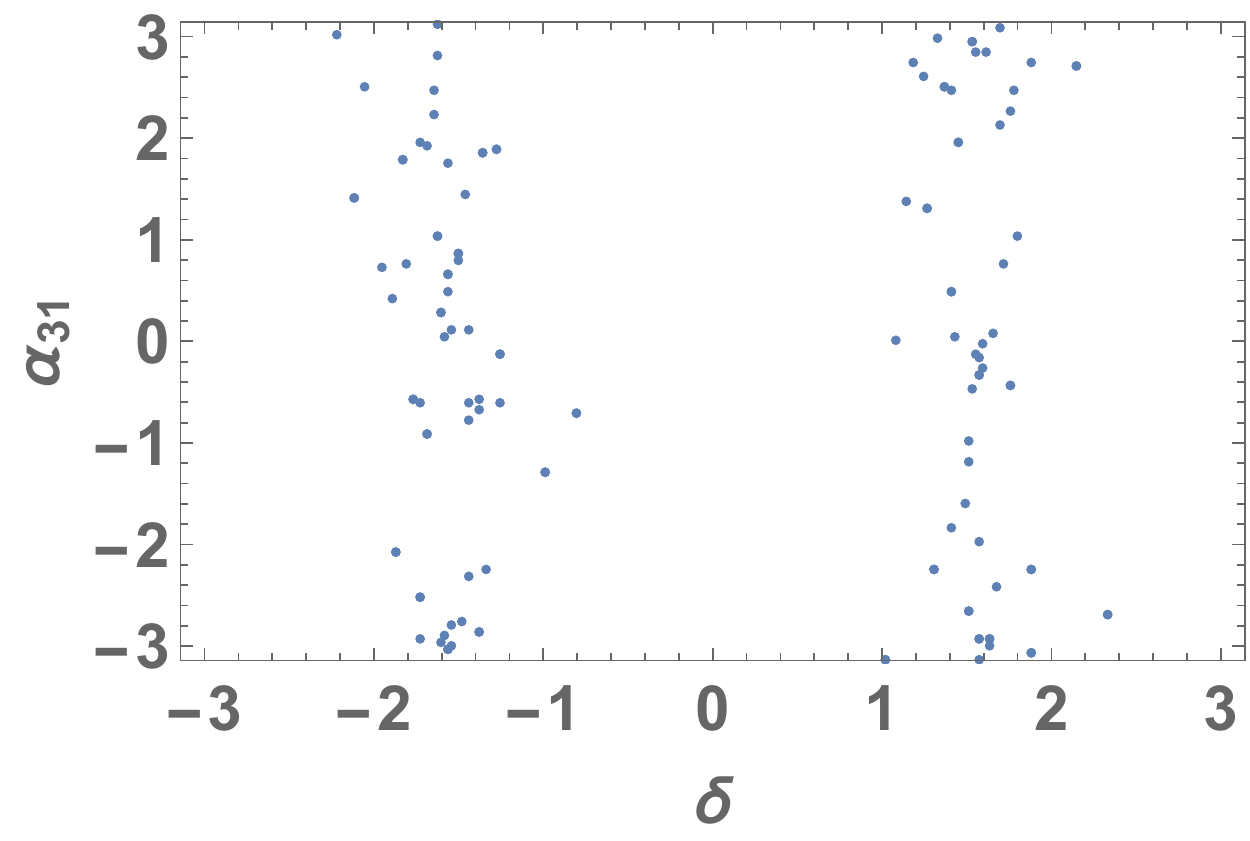}
			\caption{$\delta$ vs $\alpha_{31}$ $(\rm{I})$-C in IH}
			\label{fig.IH1CCPa31}
		\end{minipage}
	\end{tabular}
\end{figure}

\begin{figure}[h!]
	\begin{tabular}{ccc}
		\begin{minipage}{0.475\hsize}
			\includegraphics[width=\linewidth]{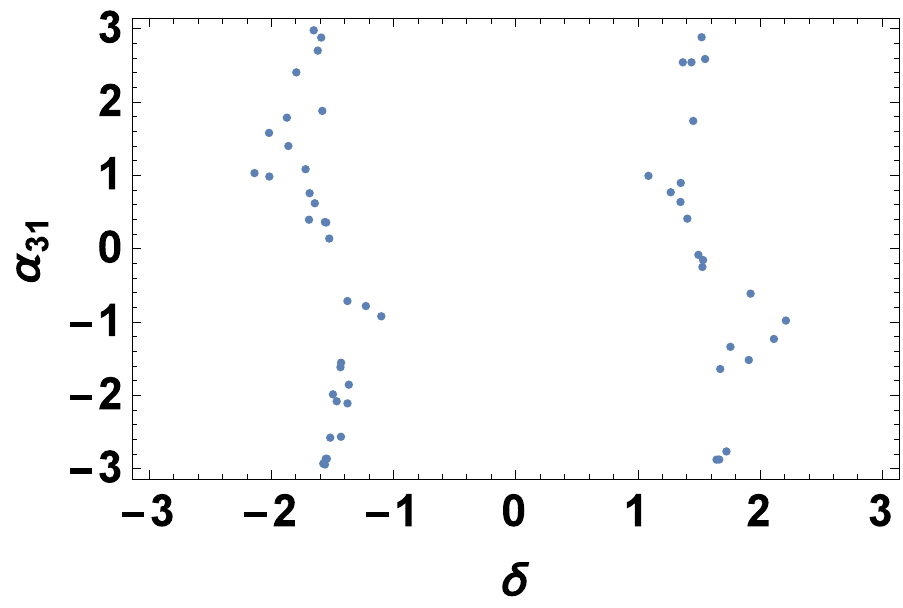}
			\caption{$\delta$ vs $\alpha_{31}$ $(\rm{I\hspace{-1pt}I\hspace{-1pt}I})$-A in IH}
			\label{fig.IH3ACPa31}
		\end{minipage}
		\phantom{=}
		\begin{minipage}{0.475\linewidth}
			\vspace{0cm}
			\includegraphics[width=0.975\linewidth]{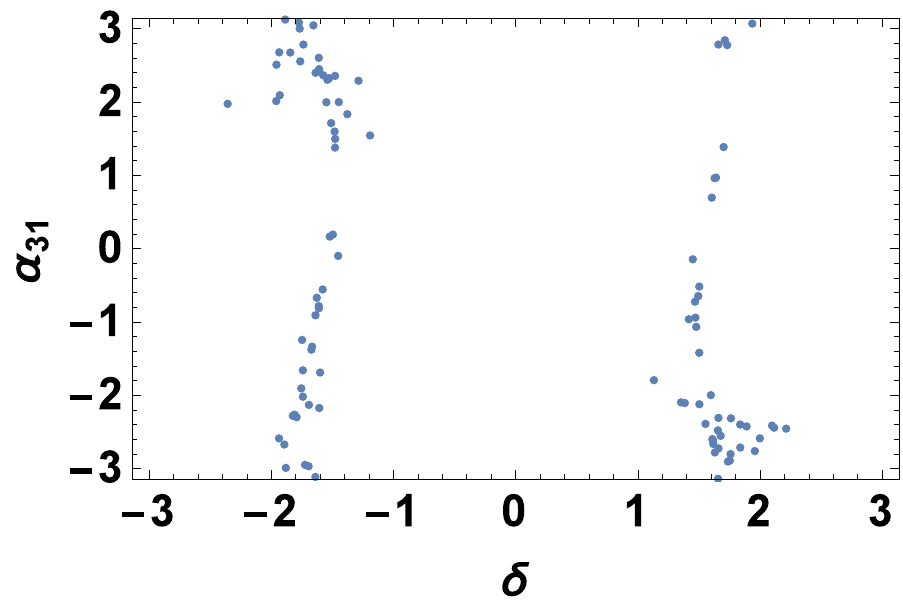}
			\caption{$\delta$ vs $\alpha_{31}$ $(\rm{I\hspace{-1pt}I\hspace{-1pt}I})$-B in IH}
			\label{fig.IH3BCPa31}
		\end{minipage}
	\end{tabular}
\end{figure}
\begin{figure}[h!]
	\begin{tabular}{ccc}
		\begin{minipage}{0.475\hsize}
			\includegraphics[width=\linewidth]{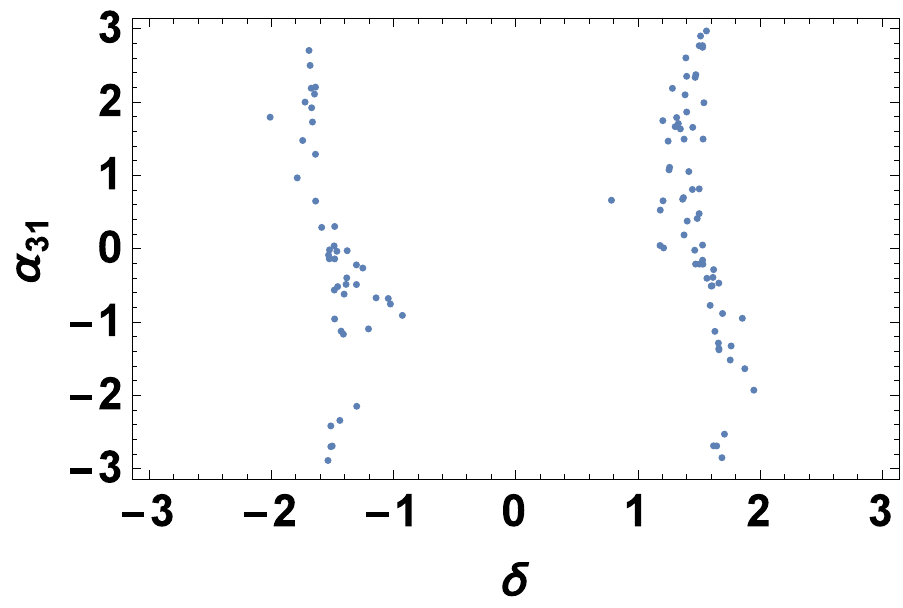}
			\caption{$\delta$ vs $\alpha_{31}$ $(\rm{I\hspace{-1pt}I\hspace{-1pt}I})$-E in IH}
			\label{fig.IH3ECPa31}
		\end{minipage}
		\phantom{=}
		\begin{minipage}{0.475\linewidth}
			\vspace{0cm}
			\includegraphics[width=0.975\linewidth]{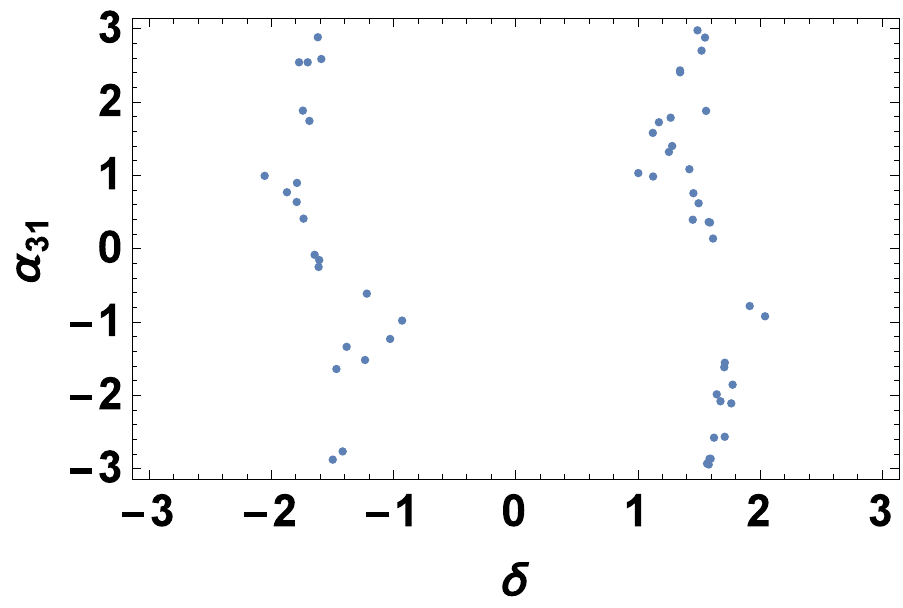}
			\caption{$\delta$ vs $\alpha_{31}$ $(\rm{I\hspace{-1pt}I\hspace{-1pt}I})$-F in IH}
			\label{fig.IH3FCPa31}
		\end{minipage}
	\end{tabular}
\end{figure}

Then, we mention the inverted hierarchical case. 
There is also a characteristic relation between the Dirac CP phase $\delta$ and one of the Majorama phases $\alpha_{21}$ in subclasses $(\rm{I})$-B,$(\rm{I})$-C, $(\rm{I\hspace{-1pt}I\hspace{-1pt}I})$-A, $(\rm{I\hspace{-1pt}I\hspace{-1pt}I})$-B, $(\rm{I\hspace{-1pt}I\hspace{-1pt}I})$-E and $(\rm{I\hspace{-1pt}I\hspace{-1pt}I})$-F.  
Their Dirac CP phase $\delta$ and the Majorama phase $\alpha_{21}$ are found in the restricted range.
In these cases, we do not find no point near $\delta=0$. 
The sign of the product of $\delta$ and $\alpha_{21}$ is determined by the subclasses.
For $(\rm{I})$-C in Fig.(\ref{fig.IH1CCPa21}),
for $(\rm{I\hspace{-1pt}I\hspace{-1pt}I})$-A in Fig.(\ref{fig.IH3ACPa21}) and for $(\rm{I\hspace{-1pt}I\hspace{-1pt}I})$-E in Fig.(\ref{fig.IH3ECPa21}), the plotted points ($\delta$, $\alpha_{21}$) are predicted only in the second and fourth quadrant.
For $(\rm{I})$-B in Fig.(\ref{fig.IH1BCPa21}), for $(\rm{I\hspace{-1pt}I\hspace{-1pt}I})$-B in Fig.(\ref{fig.IH3BCPa21}) and for $(\rm{I\hspace{-1pt}I\hspace{-1pt}I})$-F in Fig.(\ref{fig.IH3FCPa21}), they are predicted only in the first and third quadrant.
The absolute value of $\alpha_{21}$ in $(\rm{I\hspace{-1pt}I\hspace{-1pt}I})$-B in Fig.(\ref{fig.IH3FCPa21}) is bigger than those of the other three subclasses.

We find a correlation between $\delta$ and $\alpha_{21}$ in some subclasses.
On the other hand, the other Majorana phase $\alpha_{31}$ seems to have no strong correlation neither with $\delta$ nor with $\alpha_{21}$.
We show the figures $\delta$ vs $\alpha_{31}$ for the sake of comparison with $\alpha_{21}$. 
In Table \ref{table:subclassesparameters}, we show the number of figures which correspond to each subclass.
We next comment on the textures which are not consistent with the experimental data. 
We find that in normal hierarchy case, subclasses $(\rm{I})$-A,  $(\rm{I\hspace{-1pt}I\hspace{-1pt}I})$-C and  $(\rm{I\hspace{-1pt}I\hspace{-1pt}I})$-D 
are not consistent with them. 
The (2,3) and (3,2) elements of the effective mass matrix built only by these textures are equal to zero, namely $(m_{\rm{eff}})_{\mu\tau}=(m_{\rm{eff}})_{\tau\mu}=0$.
In inverted hierarchy case, only subclass $(\rm{I\hspace{-1pt}I})$-B 
does not lead to the experimental data.
In the next section, we explain the reason why these textures are not consistent with the experimental data.

\section{Analyses with hidden relations among  elements of Majorana mass matrix}

\subsection{Constraints from vanishing elements on $m_{\rm{eff}}$}
\label{sec:Constraints}

In this section, we make a discussion as for textures on the effective mass matrix.
Constructing the effective mass matrices in terms of our four-zero texture model on Dirac mass matrix, 
they are categorized into 4 cases from the view point of zero-elements-configuration on $m_{\rm{eff}}$.
Three of them correspond to the zero-off-diagonal elements of $m_{\rm{eff}}$.    
The other case corresponds to the $m_{eff}$ whose all elements are non-zero.
 \begin{equation}
 (m_{eff})_{e\mu}=(m_{eff})_{\mu e}=0,
 \end{equation}
 \begin{equation}
(m_{eff})_{e\tau}=(m_{eff})_{\tau e}=0,
 \end{equation}
 \begin{equation}
(m_{eff})_{\mu\tau}=(m_{eff})_{\tau\mu}=0,
 \end{equation}
 \begin{equation}
(m_{eff})_{\alpha\beta}\neq0 \ (\alpha,\beta=e,\mu,\tau).
 \end{equation}

Regarding our four-zero textures on Dirac mass matrix, the textures in class (I) and class $(\rm{I\hspace{-1pt}I\hspace{-1pt}I})$ automatically produce zero-elements on $m_{\rm{eff}}$, while the textures in class $(\rm{I\hspace{-1pt}I})$ lead to a $m_{\rm{eff}}$ whose all elements are non-zero.
We summarize it on the table below, which subclass of textures produces zeros on which elements of  $m_{\rm{eff}}$. 
The textures without zero-elements in $m_{\rm{eff}}$ are also shown in Table\ref{table:mefftexture}.

\begin{table}[htb]
	\begin{center}
		\caption{The textures on the effective mass matrix}
		\begin{tabular}{|l|c|c|c|c|} \hline
			& $(m_{\rm{eff}})_{e\mu}=0$ & $(m_{\rm{eff}})_{e\tau}=0$ & $(m_{\rm{eff}})_{\mu\tau}=0$ & all elements are non-zero \\ \hline 
			$m_{\rm{eff}}$ & $\left(
			\begin{array}{ccc}
			* & 0 & * \\
			0 & * & * \\
			* & * & *
			\end{array}
			\right)$ 
			& $
			\left(
			\begin{array}{ccc}
			* & * & 0 \\
			* & * & * \\
			0 & * & *
			\end{array}
			\right)$ 
			& $\left(
			\begin{array}{ccc}
			* & * & * \\
			* & * & 0 \\
			* & 0 & *
			\end{array}
			\right)$
			& $\left(
			\begin{array}{ccc}
			* & * & * \\
			* & * & * \\
			* & * & *
			\end{array}
			\right)$
			\\ \hline 
			textures & (I)-C &  (I)-B  & (I)-A & $(\rm{I\hspace{-1pt}I})$-A \\
			&$(\rm{I\hspace{-1pt}I\hspace{-1pt}I})$-B & $(\rm{I\hspace{-1pt}I\hspace{-1pt}I})$-A & $(\rm{I\hspace{-1pt}I\hspace{-1pt}I})$-C & $(\rm{I\hspace{-1pt}I})$-B \\ 
			& $(\rm{I\hspace{-1pt}I\hspace{-1pt}I})$-E & $(\rm{I\hspace{-1pt}I\hspace{-1pt}I})$-F &  $(\rm{I\hspace{-1pt}I\hspace{-1pt}I})$-D & $(\rm{I\hspace{-1pt}I})$-C \\
			\hline
		\end{tabular}
		\label{table:mefftexture}
	\end{center}
\end{table}

Which off-diagonal element of $m_{eff}$ is zero, determine the constraint among mass eigenvalues, mixing angles and CP phases.
Describing each off-diagonal elements of $m_{eff}$ in terms of $m_1$,$m_2$,$m_3$,$\theta_{12}$,$\theta_{13}$,$\theta_{23}$,$\delta$,$\alpha_{21}$, and $\alpha_{31}$,

\begin{equation}
	\begin{split}
		(m_{\rm{eff}})_{e\mu}&=\frac{1}{2}(-m_1c_{12}^2e^{i\delta}-m_2s_{12}^2e^{i(\alpha_{21}+\delta)}+m_3e^{i(\alpha_{31}-\delta)})s_{23}\sin 2\theta_{13}\\
		&+\frac{1}{2}(-m_1+m_2e^{i\alpha_{21}})c_{13}c_{23}\sin 2\theta_{12},
	\end{split}
\end{equation}

\begin{equation}
	\begin{split}
		(m_{\rm{eff}})_{e\tau}&=\frac{1}{2}(-m_1c_{12}^2e^{i\delta}-m_2s_{12}^2e^{i(\alpha_{21}+\delta)}+m_3e^{i(\alpha_{31}-\delta)})c_{23}\sin 2\theta_{13}\\
		&+\frac{1}{2}(m_1-m_2e^{i\alpha_{21}})c_{13}s_{23}\sin 2\theta_{12},
	\end{split}
\end{equation}

\begin{equation}
	\begin{split}
		(m_{\rm{eff}})_{\mu\tau}&=\frac{1}{2}\{m_1(c_{12}^2s_{13}^2e^{2i\delta}-s_{12}^2)+m_2(s_{12}^2s_{13}^2e^{i(\alpha_{21}+2\delta)}-c_{12}^2e^{i\alpha_{21}})+m_3c_{13}^2e^{i\alpha_{31}}\}\sin 2\theta_{23}\\
		&+\frac{1}{2}(m_1e^{i\delta}-m_2e^{i(\alpha_{21}+\delta)})s_{13}\cos 2\theta_{23}\sin 2\theta_{12}.
	\end{split}
\end{equation}
\[
s_{ij}=\sin\theta_{ij} , c_{ij}=\cos\theta_{ij} \
\ \ (i,j=1,2,3)
\]

If an element $(m_{\rm{eff}})_{e\mu}$ is equal to 0, both $\rm{Re}(m_{\rm{eff}})_{e\mu}$ and $\rm{Im}(m_{\rm{eff}})_{e\mu}$ must be 0.
Let us then define functions $f_1$ and $f_2$ of three mass eigenvalues, three CP phases, and mixing angles,
\begin{equation}
	f_1=-\frac{m_1c_{12}^2\cos\delta+m_2s_{12}^2\cos(\alpha_{21}+\delta)-m_3\cos(\alpha_{31}-\delta)}{m_1-m_2\cos\alpha_{21}},
	\label{f1}
\end{equation}

\begin{equation}
	f_2=\frac{m_1c_{12}^2\sin\delta+m_2s_{12}^2\sin(\alpha_{21}+\delta)-m_3\sin(\alpha_{31}-\delta)}{m_2\sin\alpha_{21}},
		\label{f2}
\end{equation}
and a function of mixing angles,
\begin{equation}
Y_1=\frac{\sin 2\theta_{12}}{2 \sin\theta_{13} \tan\theta_{23}}.
\end{equation}

The condition that $(m_{eff})_{e\mu}=0$ is identical to that
\begin{equation}
f_1=Y_1,
\label{3Dplotsf1Y1}
\end{equation}
\begin{equation}
f_2=Y_1.
\label{3Dplotsf2Y1}
\end{equation}
After substituting the experimental values of mixing angles and mass squared differences $\Delta m^2_{\rm{sol.}}$ and $\Delta m^2_{\rm{atm.}}$, $f_1$ and $f_2$ are regarded as functions of the lightest neutrino mass and three CP phases, while the value of $Y_1$ is determined. 
Eq.(\ref{3Dplotsf1Y1}) and Eq.(\ref{3Dplotsf2Y1}) imply that there are two conditions for four variables $m_1$ ($m_3$), $\delta$, $\alpha_{21}$, and $\alpha_{31}$. 

As for the condition that $(m_{eff})_{e\tau}=0$,
defining a function $Y_2$ of mixing angles
\begin{equation}
Y_2=-\frac{\sin 2\theta_{12}\tan\theta_{23}}{2 \sin\theta_{13}},
\end{equation}
two conditions among the lightest neutrino mass and CP phases are written as
\begin{equation}
f_1=Y_2,
\label{3Dplotsf1Y2}
\end{equation}
\begin{equation}
f_2=Y_2.
\label{3Dplotsf2Y2}
\end{equation}
We note that the same functions $f_1$ and $f_2$ defined in Eq.(\ref{f1}) and in Eq.(\ref{f2}) appear in Eq.(\ref{3Dplotsf1Y2}) and Eq.(\ref{3Dplotsf2Y2}).

As for the condition that $(m_{eff})_{\mu\tau}=0$,
it is written as
\begin{equation}
f_3=Y_3,
\label{3Dplotsf3Y3}
\end{equation}
\begin{equation}
f_4=Y_3.
\label{3Dplotsf4Y3}
\end{equation}
where
\begin{equation}
	f_3=\frac{m_1(c_{12}^2s_{13}^2\cos 2\delta-s_{13}^2)+m_2(s_{12}^2s_{13}^2\cos(\alpha_{21}+2\delta)-c_{12}^2\cos\alpha_{21})+m_3c_{13}^2\cos\alpha_{31}}{m_1\cos\delta-m_2\cos(\alpha_{21}+\delta)},
\end{equation}

\begin{equation}
	f_4=\frac{m_1c_{12}^2s_{13}^2\sin 2\delta+m_2(s_{12}^2s_{13}^2\sin(\alpha_{21}+2\delta)-c_{12}^2\sin\alpha_{21})+m_3c_{13}^2\sin\alpha_{31}}{m_1\sin\delta-m_2\sin(\alpha_{21}+\delta)},
\end{equation}
and
\begin{equation}
	Y_3=-\frac{\sin 2\theta_{12}\sin\theta_{13}}{\tan 2\theta_{23}}.
\end{equation}

Also in the case of $m_{eff}$ whose all elements are not zero, we can find conditions among the lightest neutrino mass and three CP phases. Such $m_{eff}$ is made of a Dirac mass matrix with textures in class$(\rm{I\hspace{-1pt}I})$.
By taking an example from the subclass$(\rm{I\hspace{-1pt}I})$-B, a Dirac mass matrix of the texture
\begin{equation}
	\left(
	\begin{array}{ccc}
		\sin\theta_1\cos\theta_2 & 0 & 0 \\
		\sin\theta_1\sin\theta_2 & e^{i\phi_1} & 0 \\
		\cos\theta_1 & 0 & e^{i\phi_2}
	\end{array}
	\right)
\end{equation}
produces the effective mass matrix
\begin{equation}
	m_{eff}=-\left(
	\begin{array}{ccc}
		X_1\sin^2\theta_1\cos^2\theta_2 & \frac{1}{2}X_1\sin^2\theta_1\sin 2\theta_2 & \frac{1}{2}X_1\sin 2\theta_1\cos\theta_2 \\
		\frac{1}{2}X_1\sin^2\theta_1\sin 2\theta_2 & X_1\sin^2\theta_1\sin^2\theta_2+X_2e^{2i\phi_1} & \frac{1}{2}X_1\sin 2\theta_1\sin\theta_2 \\
		\frac{1}{2}X_1\sin 2\theta_1\cos\theta_2 & \frac{1}{2}X_1\sin 2\theta_1\sin\theta_2 & X_3e^{2i\phi_2}
	\end{array}
	\right).
	\label{class2example}
\end{equation}
Let us consider a quantity independent of the phase redefinition for charged lepton's flavor basis,
\begin{equation}
	\frac{(m_{\rm{eff}})_{ee}(m_{\rm{eff}})_{\mu\tau}}{(m_{\rm{eff}})_{e\mu}(m_{\rm{eff}})_{e\tau}}.
\end{equation}
If $m_{\rm{eff}}$ is given as  Eq.(\ref{class2example}), the product equals to 1.
\begin{equation}
\frac{(m_{\rm{eff}})_{ee}(m_{\rm{eff}})_{\mu\tau}}{(m_{\rm{eff}})_{e\mu}(m_{\rm{eff}})_{e\tau}}=1
\label{invariant1}
\end{equation}
We generalize this condition to the all textures in class$(\rm{I\hspace{-1pt}I})$. Setting $\alpha$ as the row component which has only one non-zero elements on the Dirac mass matrix in a texture of class$(\rm{I\hspace{-1pt}I})$, and $\beta$ and $\gamma$ as the other row components which have two non-zero elements,   
the generalization of Eq.(\ref{invariant1}) is
\begin{equation}
	\frac{(m_{\rm{eff}})_{\alpha\alpha}(m_{\rm{eff}})_{\beta\gamma}}{(m_{\rm{eff}})_{\alpha\beta}(m_{\rm{eff}})_{\alpha\gamma}}=1.
	\label{invariant}
\end{equation}
\[
(\alpha, \beta ,\gamma=e,\mu,\tau \ \ \ \  \alpha\neq\beta\neq\gamma) 
\]
Eq.(\ref{invariant}) is symmetric with respect to the exchange between $\beta$ and $\gamma$.
For the case of the subclass $(\rm{I\hspace{-1pt}I})$-A, $\alpha$ stands for $\tau$. 
For the subclass $(\rm{I\hspace{-1pt}I})$-B and $(\rm{I\hspace{-1pt}I})$-C, $\alpha$ stands for $e$ and $\mu$ respectively.
Rewriting the elements of $m_{eff}$ in terms of mass eigenvalues, mixing angles, and CP phases, we obtain two conditions among them.
\begin{equation}
	\rm{Re}\Bigl{(}\frac{(m_{eff})_{\alpha\alpha}(m_{eff})_{\beta\gamma}}{(m_{eff})_{\alpha\beta}(m_{eff})_{\alpha\gamma}}\Bigr{)}=1
	\label{Re1}
\end{equation}
\begin{equation}
	\rm{Im}\Bigl{(}\frac{(m_{eff})_{\alpha\alpha}(m_{eff})_{\beta\gamma}}{(m_{eff})_{\alpha\beta}(m_{eff})_{\alpha\gamma}}\Bigr{)}=0
		\label{Im1}
\end{equation}
Eq.(\ref{invariant}) is also rewritten as 
\begin{equation}
\rm{Re}\{(m_{eff})_{\alpha\alpha}(m_{eff})_{\beta\gamma}\}=\rm{Re}\{(m_{eff})_{\alpha\beta}(m_{eff})_{\alpha\gamma}\},
	\label{Re2}
\end{equation} 
\begin{equation}
\rm{Im}\{(m_{eff})_{\alpha\alpha}(m_{eff})_{\beta\gamma}\}=\rm{Im}\{(m_{eff})_{\alpha\beta}(m_{eff})_{\alpha\gamma}\}.
	\label{Im2}
\end{equation} 
Each condition, Eq.(\ref{Re2}) or Eq.(\ref{Im2}), is  dependent on phase redefinition. However, two conditions all together are
equivalent to Eq.(\ref{invariant}).

\subsection{Explanations for the results of numerical analysis}

In this subsection, we give detailed explanations for the correlations among CP phases in Sec.\ref{sec:results} from a view point of the conditions discussed in Sec.\ref{sec:Constraints}.
The constraints obtained in Sec.\ref{sec:Constraints} describe the relations as for CP phases $\delta$, $\alpha_{21}$ and $\alpha_{31}$. 
We first show the relations among CP phases from $(m_{eff})_{e\mu}=0$ in IH case and from $(m_{eff})_{e\tau}=0$ in IH case, in Figs.(\ref{3D_IH_emu_01})-(\ref{3D_IH_etau_10}), which lead to a good explanation for the scatter plots in Sec.\ref{sec:results} of the corresponding four-zero texture.
$(m_{eff})_{\mu\tau}=0$ in NH case and the subclass $(\rm{I\hspace{-1pt}I})$-B in IH case do not have allowed region under the constraints. It is shown in Figs.(\ref{3D_NH_mutau_01})-(\ref{3D_IH_2B_10}). These two unfavorable cases correspond to the four-zero textures which produce no scatter plots in Sec.\ref{sec:results}.
We also mention the constraints from $(m_{eff})_{e\tau}=0$ in NH case.  Although it seems that there is less obvious relation between them in comparison with the former cases, it is not inconsistent with the results of scatter plots in Sec.\ref{sec:results}.
We show the obtained relations in Figs.(\ref{3D_NH_etau_01})-(\ref{3D_NH_etau_10}) and clarify the reason.

\subsubsection{$(m_{eff})_{e\mu}=0$ in IH case}
\label{sec:emu_zero}

We first investigate the textures of subclasses $(\rm{I})$-C, $(\rm{I\hspace{-1pt}I\hspace{-1pt}I})$-B and $(\rm{I\hspace{-1pt}I\hspace{-1pt}I})$-E, which produce 0 on $(m_{eff})_{e\mu}$, in inverted hierarchical case.
The results of numerical analysis for these textures in IH case are shown in Fig.(\ref{fig.IH1CCPa21}), Fig.(\ref{fig.IH3BCPa21}),  Fig.(\ref{fig.IH3FCPa21}), Fig.(\ref{fig.IH1CCPa31}), Fig.(\ref{fig.IH3BCPa31}) and Fig.(\ref{fig.IH3FCPa31}).
Let us compare these plotted points on the two-dimensional planes $(\delta,\alpha_{21})$ and $(\delta,\alpha_{31})$ with constraints among $\delta$, $\alpha_{21}$ and $\alpha_{31}$ under the condition Eq.(\ref{3Dplotsf1Y1}) and Eq.(\ref{3Dplotsf2Y1}). 
In IH case, two mass eigenvalues $m_1$ and $m_2$ are written as the functions of the lightest neutrino mass $m_3$ by using experimental values of the mass squared differences $\Delta m^2_{\rm{sol.}}$ and $\Delta m^2_{\rm{atm.}}$.
After substituting the values of three mixing angles  and determining the value of $m_3$, $f_1$ and $f_2$ are functions of three CP phases, and $Y_1$ has an unique value.
Eq.(\ref{3Dplotsf1Y1}) and Eq.(\ref{3Dplotsf2Y1}) imply two independent conditions for three CP phases $\delta$, $\alpha_{21}$ and $\alpha_{31}$.

For some given values of $m_3$, we draw two curved surfaces
which are described by Eq.(\ref{3Dplotsf1Y1}) and Eq.(\ref{3Dplotsf2Y1}) on a 3-dimensional space spanned by $\delta$, $\alpha_{21}$ and $\alpha_{31}$.
We substitute values of $m_3$ by dividing it into several intervals from 0 \rm{eV} to its upper bound 0.046 \rm{eV}.
The points on the nodal line of two curved surfaces indicate the possible ranges of CP phases under the condition of the vanishing element $(m_{\rm{eff}})_{e\mu}=0$. 
We show three examples of such 3-dimensional plots on $(\delta,\alpha_{21},\alpha_{31})$ space,
by taking $m_3$ as 0 \rm{eV}, the half-value of its upper bound 0.023 \rm{eV} and its upper bound 0.046 \rm{eV}.

\begin{figure}[h!]
	\begin{tabular}{ccc}
		\begin{minipage}{0.3\hsize}
			\includegraphics[width=\linewidth]{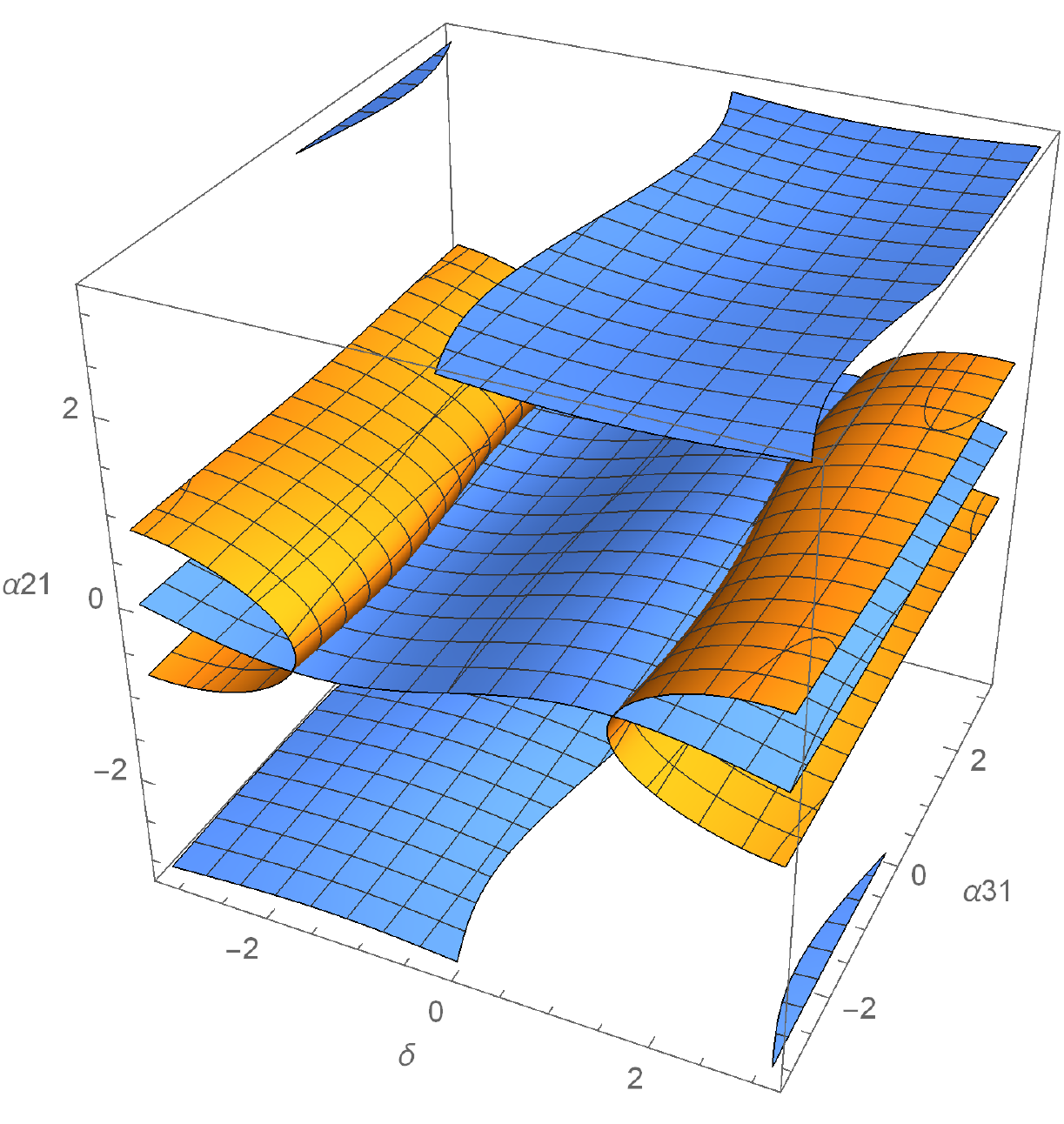}
			\caption{Contour plots on $(\delta,\alpha_{21},\alpha_{31})$ under the condition $(m_{\rm{eff}})_{e\mu}=0$ in IH, where $m_3$=0 \rm{eV}}
			\label{3D_IH_emu_01}
		\end{minipage}
		\phantom{=}
		\begin{minipage}{0.3\linewidth}
			\vspace{0cm}
			\includegraphics[width=0.975\linewidth]{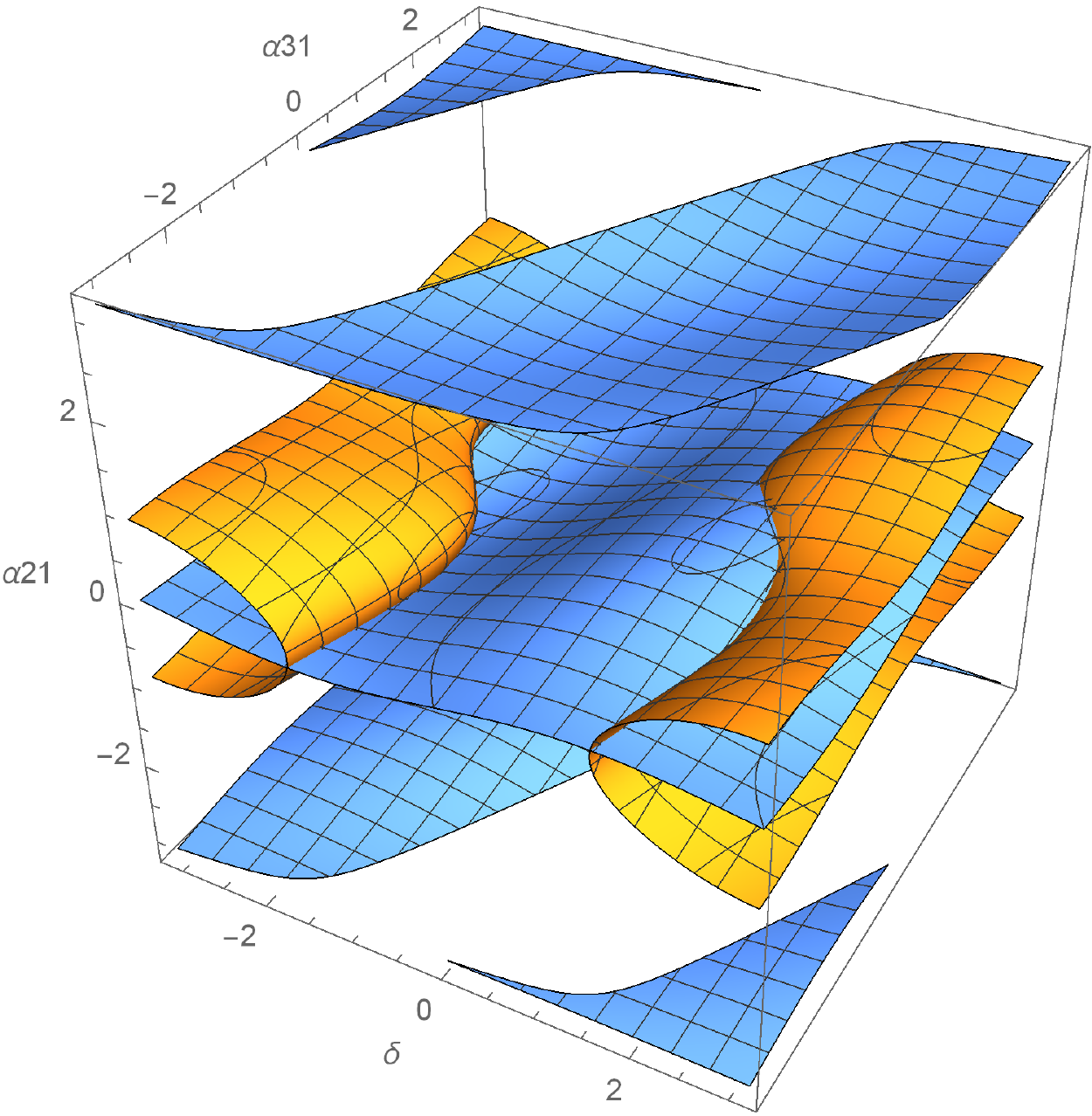}
			\caption{Contour plots on $(\delta,\alpha_{21},\alpha_{31})$ under the condition $(m_{\rm{eff}})_{e\mu}=0$ in IH, where $m_3$=0.023 [\rm{eV}]}
			\label{3D_IH_emu_05}
		\end{minipage}
		\phantom{=}
		\begin{minipage}{0.3\linewidth}
			\vspace{0cm}
			\includegraphics[width=0.975\linewidth]{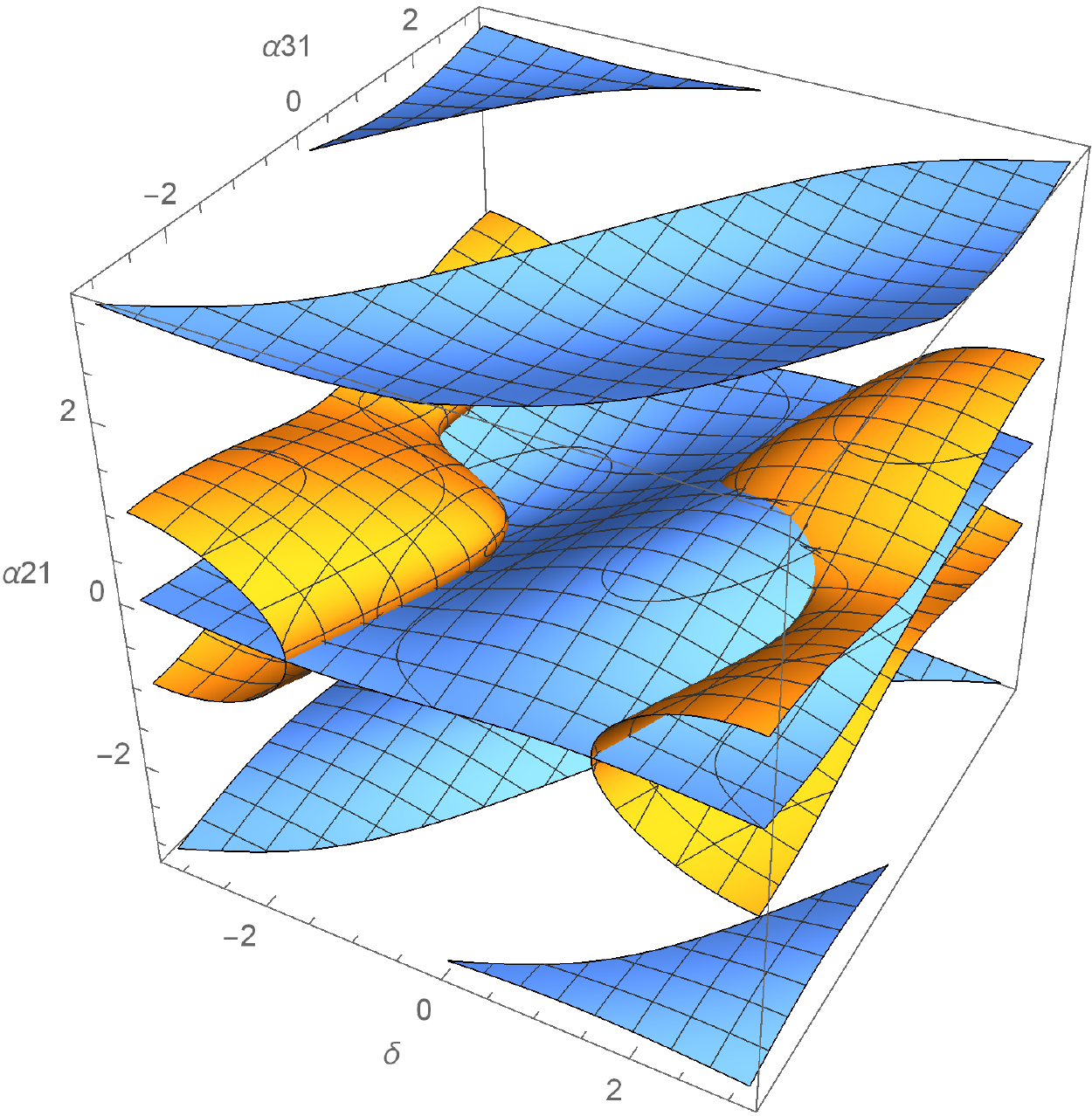}
			\caption{Contour plots on $(\delta,\alpha_{21},\alpha_{31})$ under the condition $(m_{\rm{eff}})_{e\mu}=0$ in IH, where $m_3$=0.046 \rm{eV}}
			\label{3D_IH_emu_10}
		\end{minipage}
	\end{tabular}
\end{figure}

The yellow and blue curved surfaces in Figs.(\ref{3D_IH_emu_01})-(\ref{3D_IH_emu_10}) are described by Eq.(\ref{3Dplotsf1Y1}) and Eq.(\ref{3Dplotsf2Y1}) respectively.
Two surfaces vary smoothly depending on the value of the lightest neutrino mass $m_3$.
Focusing on $(\delta,\alpha_{21})$ planes, it can be seen that the features which we have discussed in Fig.(\ref{fig.IH1CCPa21}), Fig.(\ref{fig.IH3BCPa21}) and Fig.(\ref{fig.IH3FCPa21}) are reflected in Figs.(\ref{3D_IH_emu_01})-(\ref{3D_IH_emu_10}).
By contrast, they show that $\alpha_{31}$ ranges from $-\pi$ to $\pi$ without strong dependence on $\delta$ nor $\alpha_{21}$.
This result corresponds to Fig.(\ref{fig.IH1CCPa31}), Fig.(\ref{fig.IH3BCPa31}) and Fig.(\ref{fig.IH3FCPa31}).    

\subsubsection{$(m_{eff})_{e\tau}=0$ in IH case}

\begin{figure}[h!]
	\begin{tabular}{ccc}
		\begin{minipage}{0.3\hsize}
			\includegraphics[width=\linewidth]{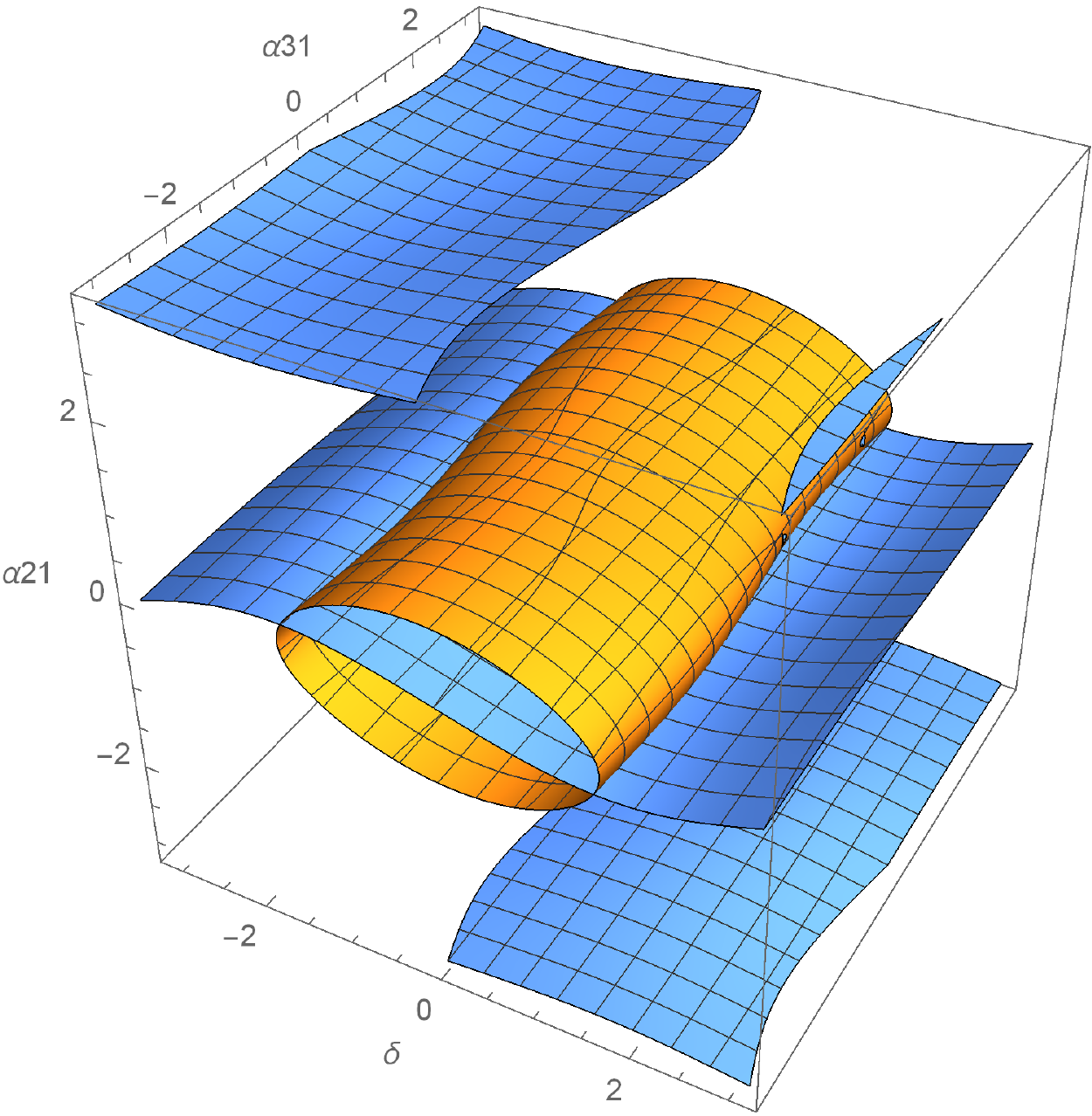}
			\caption{Contour plots on $(\delta,\alpha_{21},\alpha_{31})$ under the condition $(m_{\rm{eff}})_{e\tau}=0$ in IH, where $m_3$=0 \rm{eV}}
			\label{3D_IH_etau_01}
		\end{minipage}
		\phantom{=}
		\begin{minipage}{0.3\linewidth}
			\vspace{0cm}
			\includegraphics[width=0.975\linewidth]{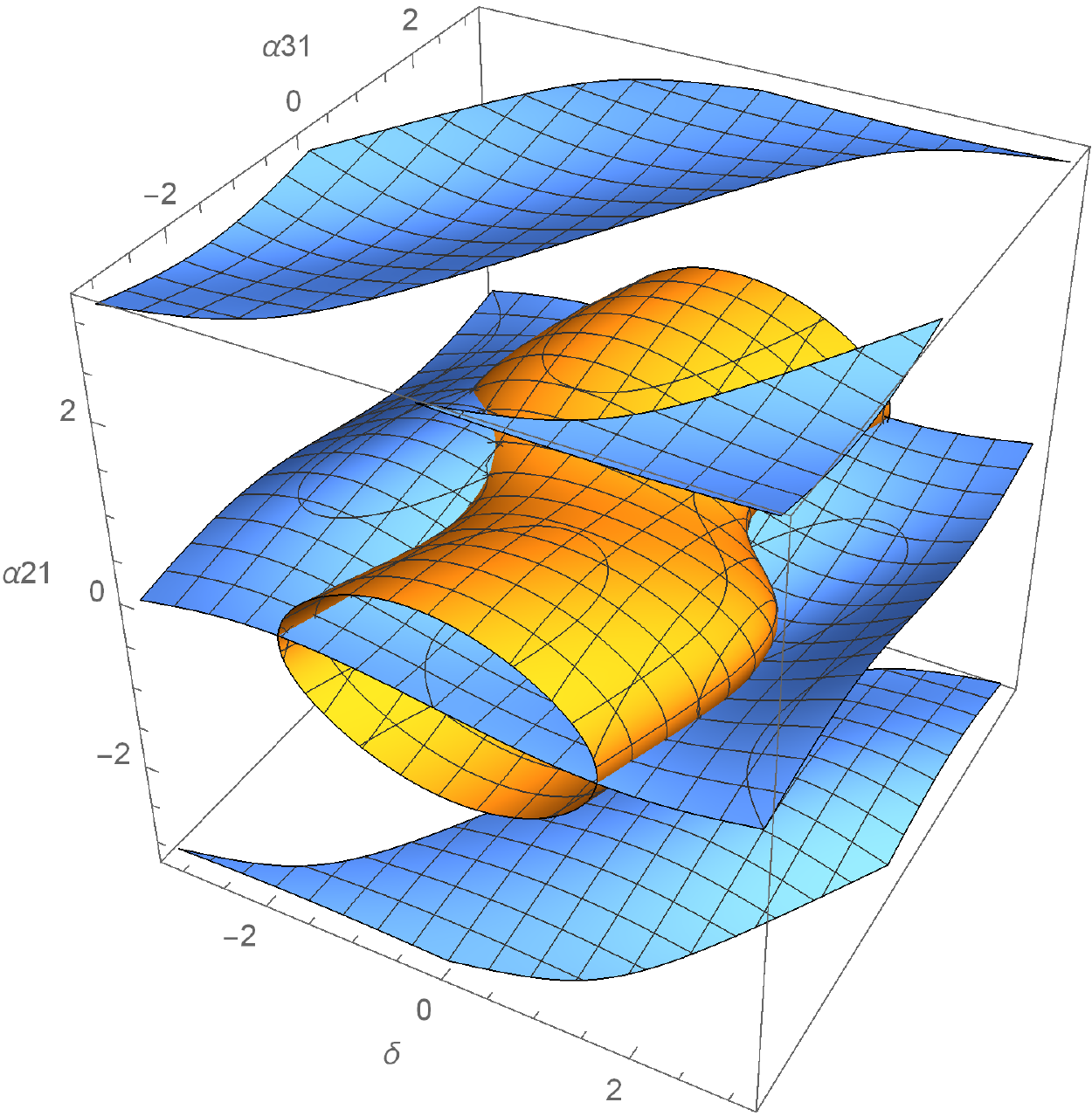}
			\caption{Contour plots on $(\delta,\alpha_{21},\alpha_{31})$ under the condition $(m_{\rm{eff}})_{e\tau}=0$ in IH, where $m_3$=0.023 \rm{eV}}
			\label{3D_IH_etau_05}
		\end{minipage}
			\phantom{=}
			\begin{minipage}{0.3\linewidth}
				\vspace{0cm}
				\includegraphics[width=0.975\linewidth]{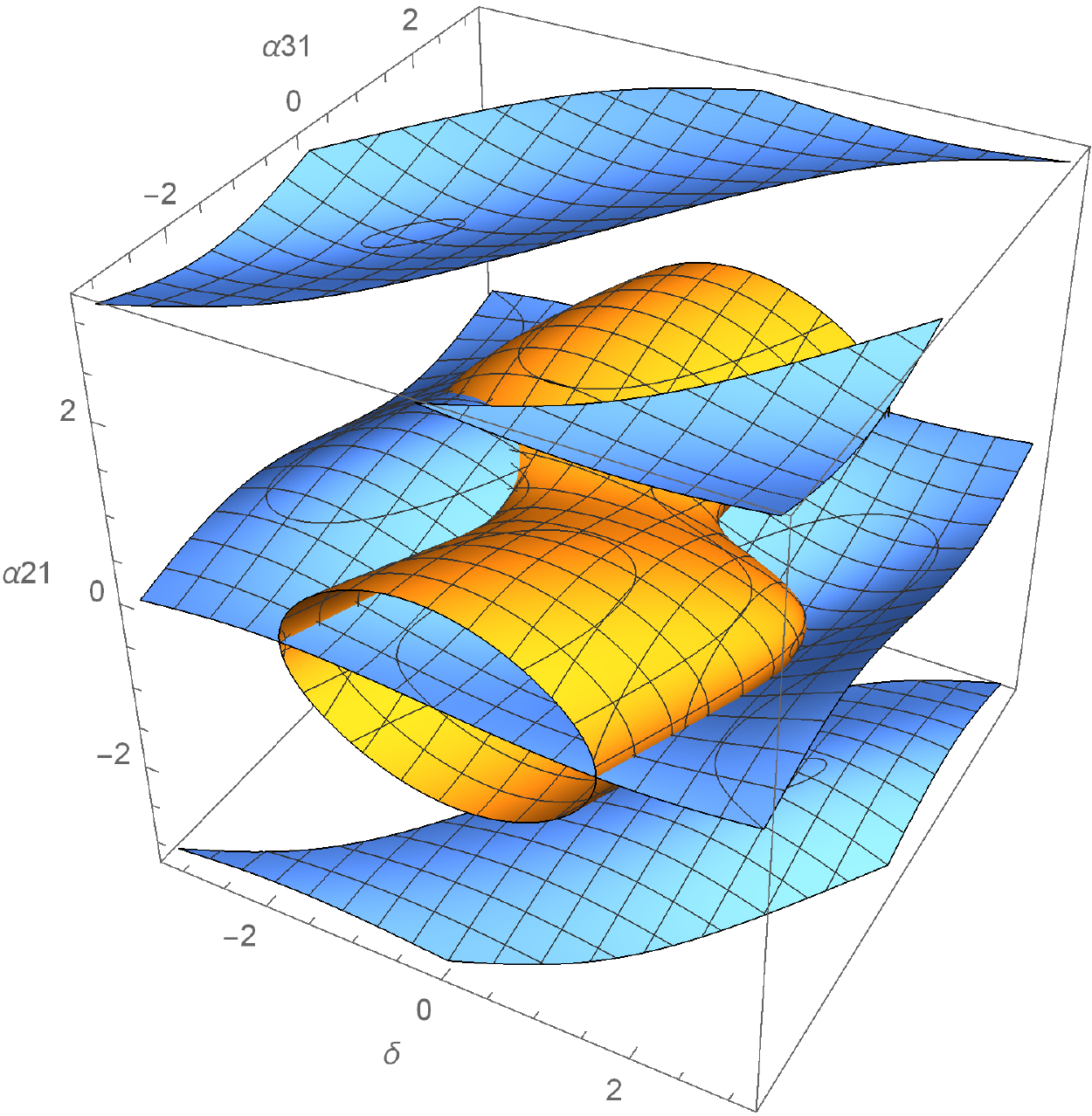}
				\caption{Contour plots on $(\delta,\alpha_{21},\alpha_{31})$ under the condition $(m_{\rm{eff}})_{e\tau}=0$ in IH, where $m_3$=0.046 \rm{eV}}
				\label{3D_IH_etau_10}
			\end{minipage}
	\end{tabular}
\end{figure}

We apply the discussion in Sec.\ref{sec:emu_zero} to another element $(m_{\rm{eff}})_{e\tau}$ of the effective mass matrix.
The results of numerical analysis for textures $(\rm{I})$-B, $(\rm{I\hspace{-1pt}I\hspace{-1pt}I})$-A and $(\rm{I\hspace{-1pt}I\hspace{-1pt}I})$-F are shown in Fig.(\ref{fig.IH1BCPa21}), Fig.(\ref{fig.IH3ACPa21}), Fig.(\ref{fig.IH3ECPa21}),Fig.(\ref{fig.IH1BCPa31}), Fig.(\ref{fig.IH3ACPa31}) and Fig.(\ref{fig.IH3ECPa31}).
These textures produce that $(m_{\rm{eff}})_{e\tau}$ equals 0.
In this case, three CP phases and the lightest neutrino mass follow the conditions Eq.(\ref{3Dplotsf1Y2}) and Eq.(\ref{3Dplotsf2Y2}).
By imposing the IH case and using the experimental values, we obtain two constraints for three CP phases.
Drawing two curved surfaces described by Eq.(\ref{3Dplotsf1Y2}) and Eq.(\ref{3Dplotsf2Y2}) on $(\delta,\alpha_{21},\alpha_{31})$ space for several values of $m_3$, we find nodal lines of them.
We show three examples in the cases of 0, the half-value of the upper bound and the upper bound of $m_3$. 
The yellow and blue curved surfaces in Figs.(\ref{3D_IH_etau_01})-(\ref{3D_IH_etau_10}) stand for Eq.(\ref{3Dplotsf1Y2}) and Eq.(\ref{3Dplotsf2Y2}) respectively.  
The allowed values on $(\delta,\alpha_{21},\alpha_{31})$ space from two curved surfaces show the similar features as Figs.(\ref{3D_IH_emu_01})-(\ref{3D_IH_emu_10}).

\subsubsection{$(m_{eff})_{\mu\tau}=0$ in NH case}

\begin{figure}[h!]
	\begin{tabular}{ccc}
		\begin{minipage}{0.3\hsize}
			\includegraphics[width=\linewidth]{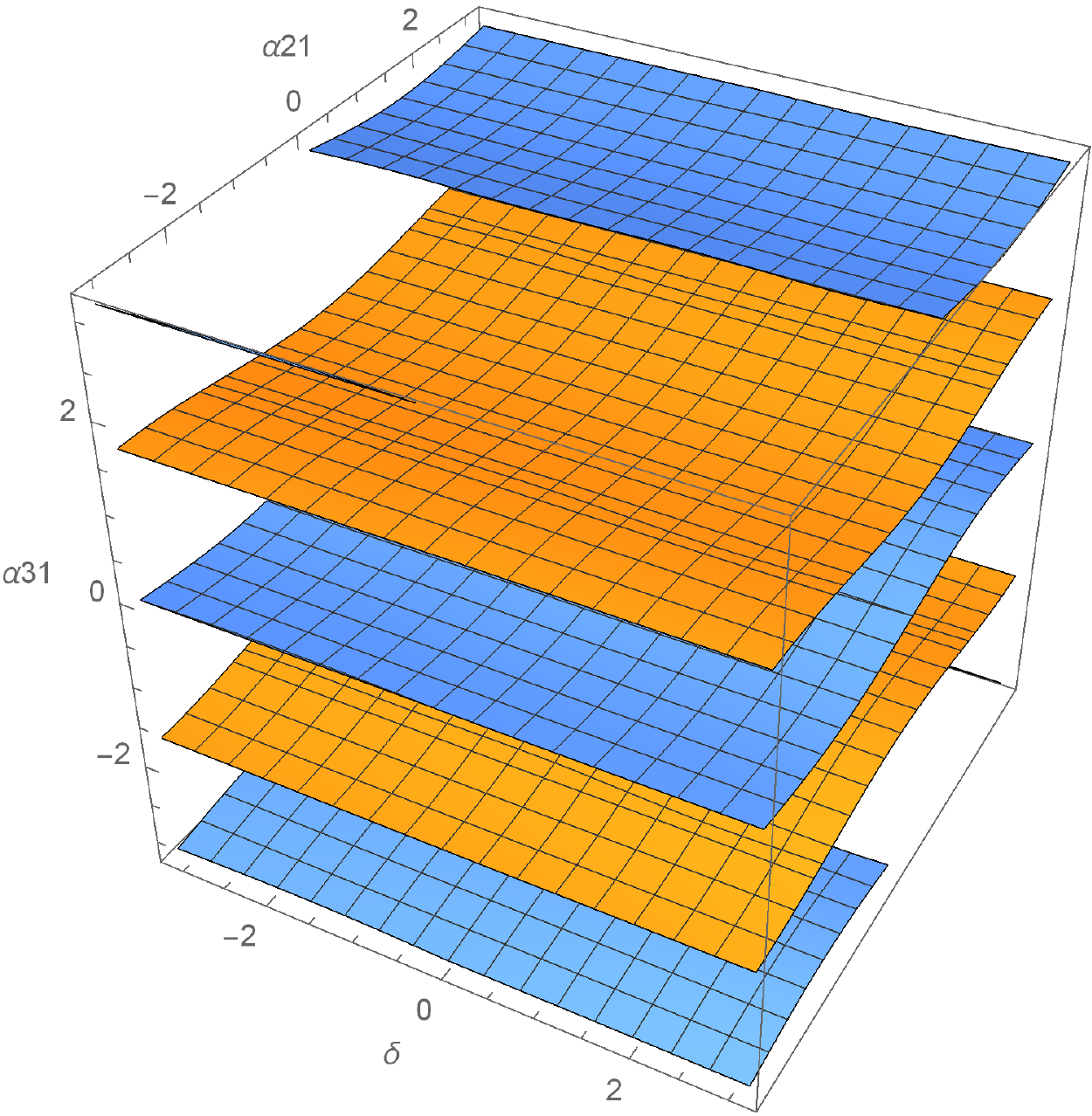}
			\caption{Contour plots on $(\delta,\alpha_{21},\alpha_{31})$ under the condition $(m_{\rm{eff}})_{\mu\tau}=0$ in NH, where $m_1$=0 \rm{eV}}
			\label{3D_NH_mutau_01}
		\end{minipage}
		\phantom{=}
		\begin{minipage}{0.3\linewidth}
			\vspace{0cm}
			\includegraphics[width=0.975\linewidth]{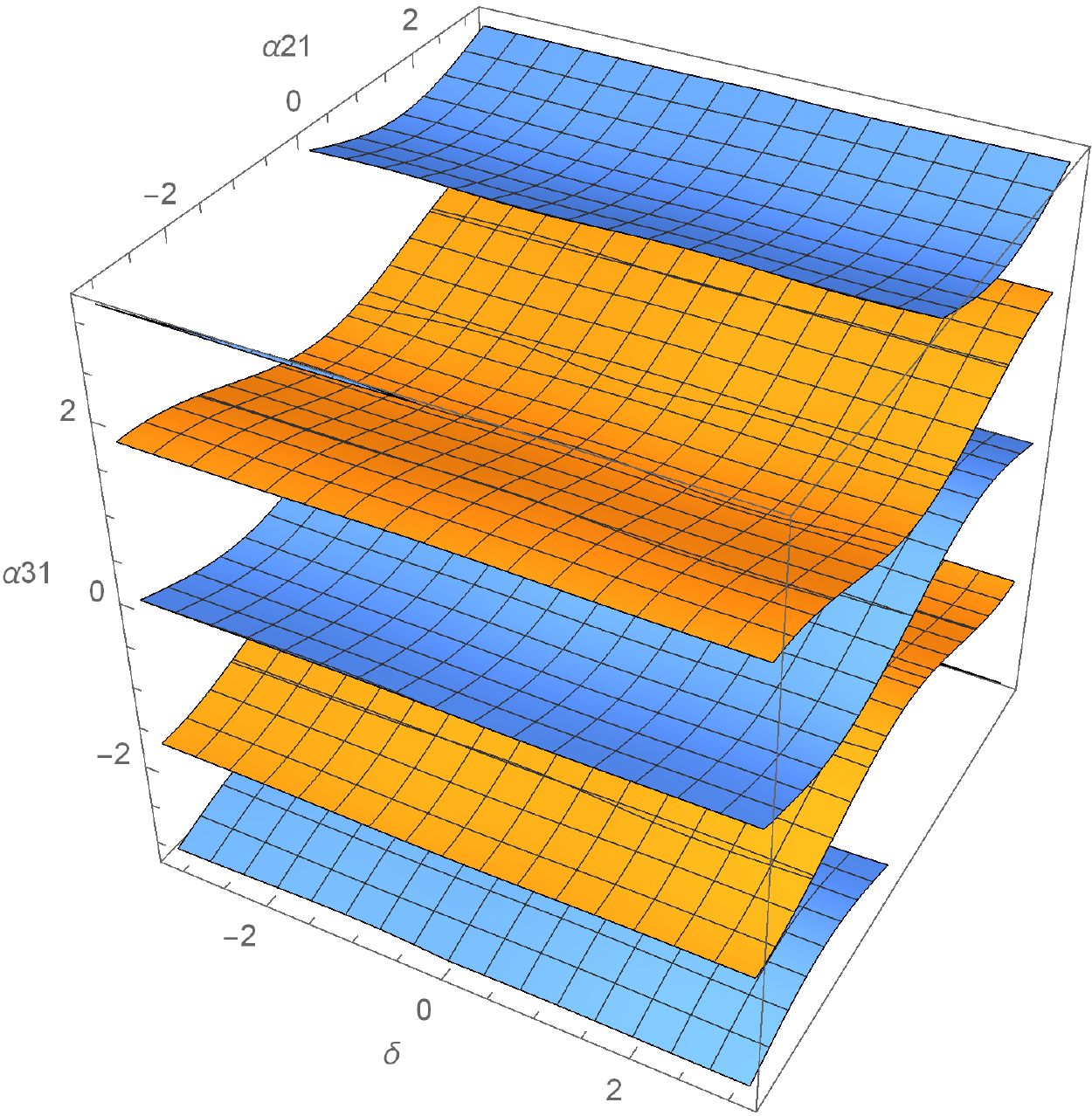}
			\caption{Contour plots on $(\delta,\alpha_{21},\alpha_{31})$ under the condition $(m_{\rm{eff}})_{\mu\tau}=0$ in NH, where $m_1$=0.023 \rm{eV}}
			\label{3D_NH_mutau_05}
		\end{minipage}
		\phantom{=}
		\begin{minipage}{0.3\linewidth}
			\vspace{0cm}
			\includegraphics[width=0.975\linewidth]{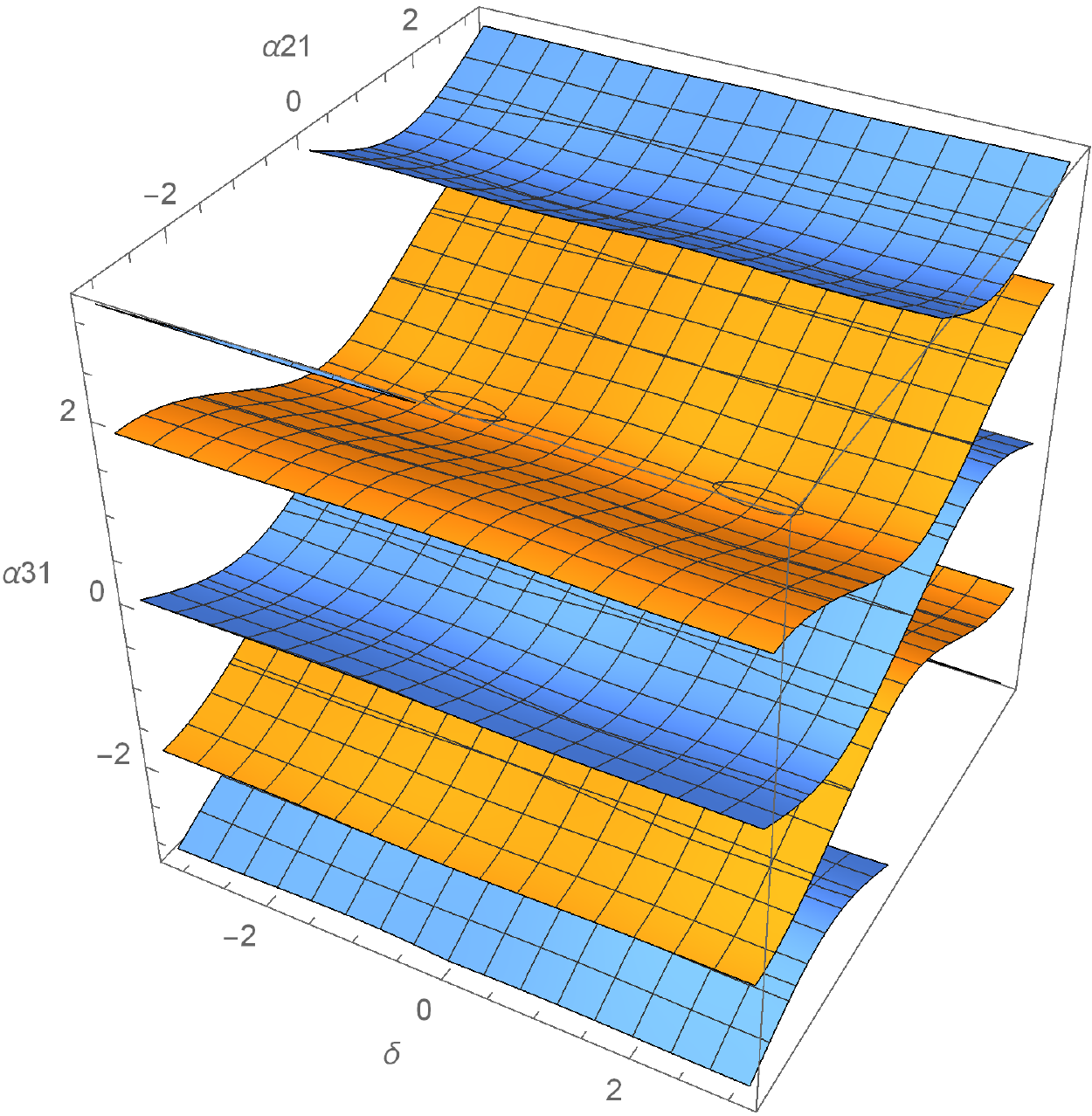}
			\caption{Contour plots on $(\delta,\alpha_{21},\alpha_{31})$ under the condition $(m_{\rm{eff}})_{\mu\tau}=0$ in NH, where $m_1$=0.046 \rm{eV}}
			\label{3D_NH_mutau_10}
		\end{minipage}
	\end{tabular}
\end{figure}

We have mentioned that subclasses $(\rm{I})$-A,  $(\rm{I\hspace{-1pt}I\hspace{-1pt}I})$-C and  $(\rm{I\hspace{-1pt}I\hspace{-1pt}I})$-D, which produce 0 on $(m_{\rm{eff}})_{\mu\tau}$, were not consistent with the experimental data in normal hierarchy case.
It can be explained by the conditions Eq.(\ref{3Dplotsf3Y3}) and Eq.(\ref{3Dplotsf4Y3}). 
In NH case, two mass eigenvalues $m_2$ and $m_3$ are written as the functions of the lightest neutrino mass $m_1$.
By using the experimental values of mixing angles and determining the value of $m_1$, Eq.(\ref{3Dplotsf3Y3}) and Eq.(\ref{3Dplotsf4Y3}) can be regarded as two independent conditions for three CP phases.
We can draw them on $(\delta,\alpha_{21},\alpha_{31})$ space.
The value of $m_1$ is taken similarly by several intervals from 0 \rm{eV} to 0.046 \rm{eV}.
Two curved surfaces vary smoothly depending on $m_1$.
As the value of $m_1$ is increased, these surfaces are approaching to each other.
However, they do not intersect within the upper limit of $m_1$.  
We show three examples for $m_1=0$, $0.023$ and $0.046$ eV.
The yellow and blue curved surfaces in Figs.(\ref{3D_NH_mutau_01})-(\ref{3D_NH_mutau_10}) stand for Eq.(\ref{3Dplotsf3Y3}) and Eq.(\ref{3Dplotsf4Y3}) respectively.
Therefore, the textures whose  $(m_{\rm{eff}})_{\mu\tau}$ is equal to zero do not reproduce the experimentally measured mixing angles and mass eigenvalues in NH case.

\begin{figure}[h!]
	\begin{tabular}{ccc}
		\begin{minipage}{0.3\hsize}
			\includegraphics[width=\linewidth]{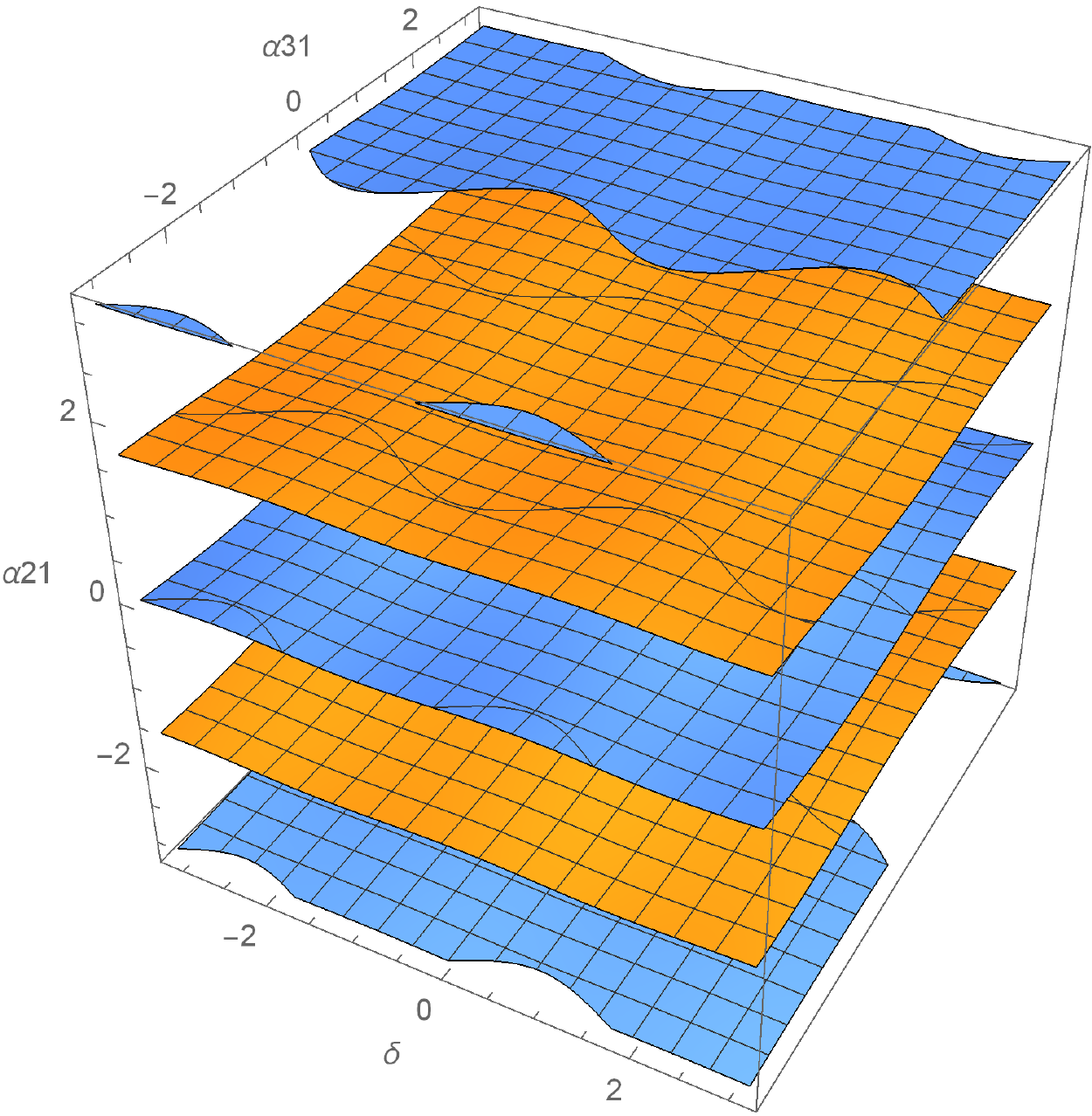}
			\caption{Contour plots on $(\delta,\alpha_{21},\alpha_{31})$ for subclass $(\rm{I\hspace{-1pt}I})$-B in IH, where $m_3$=0 \rm{eV}}
			\label{3D_IH_2B_01}
		\end{minipage}
		\phantom{=}
		\begin{minipage}{0.3\linewidth}
			\vspace{0cm}
			\includegraphics[width=0.975\linewidth]{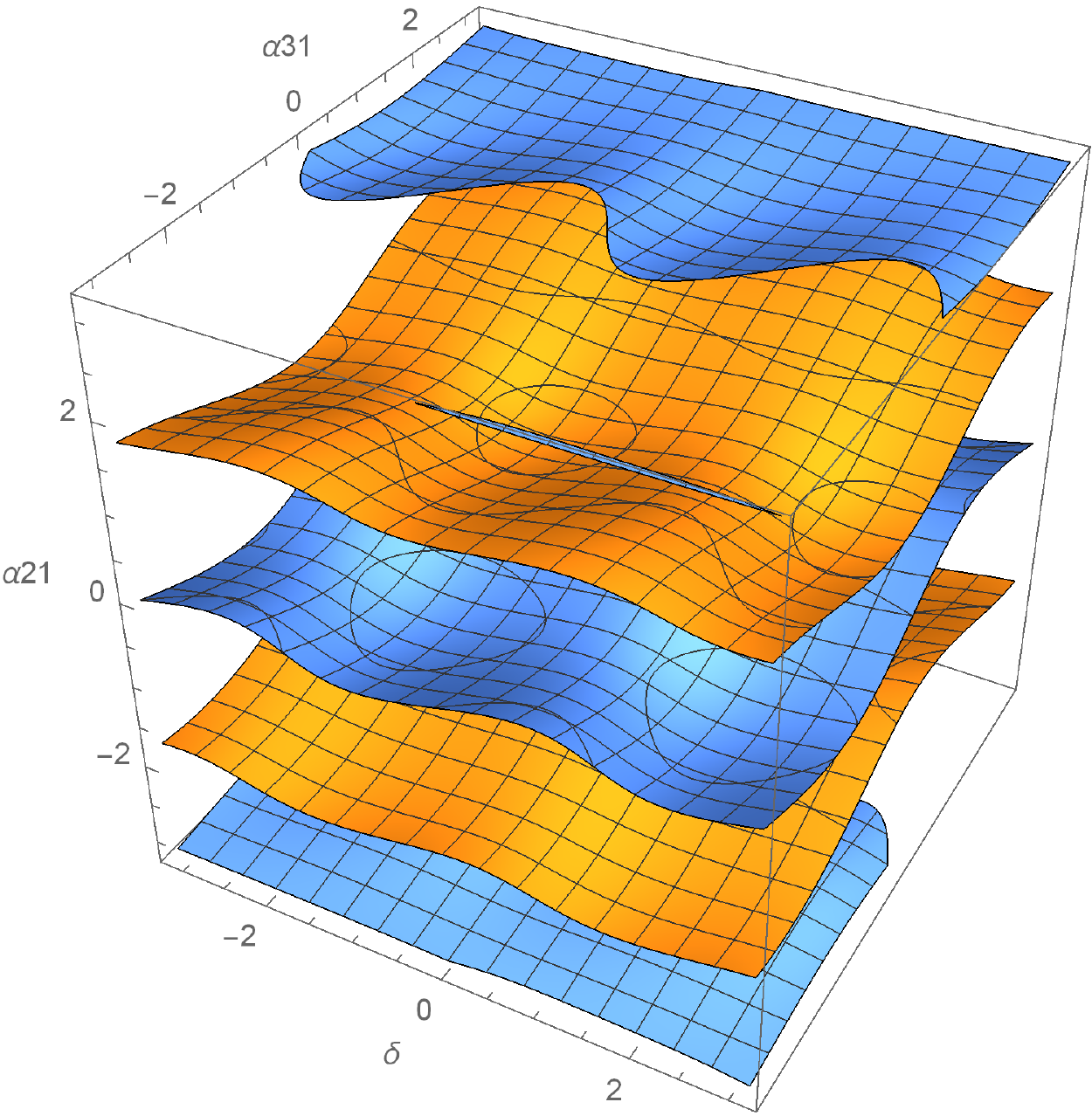}
			\caption{Contour plots on $(\delta,\alpha_{21},\alpha_{31})$ for subclass $(\rm{I\hspace{-1pt}I})$-B in IH, where $m_3$=0.023 \rm{eV}}
			\label{3D_IH_2B_05}
		\end{minipage}
		\phantom{=}
		\begin{minipage}{0.3\linewidth}
			\vspace{0cm}
			\includegraphics[width=0.975\linewidth]{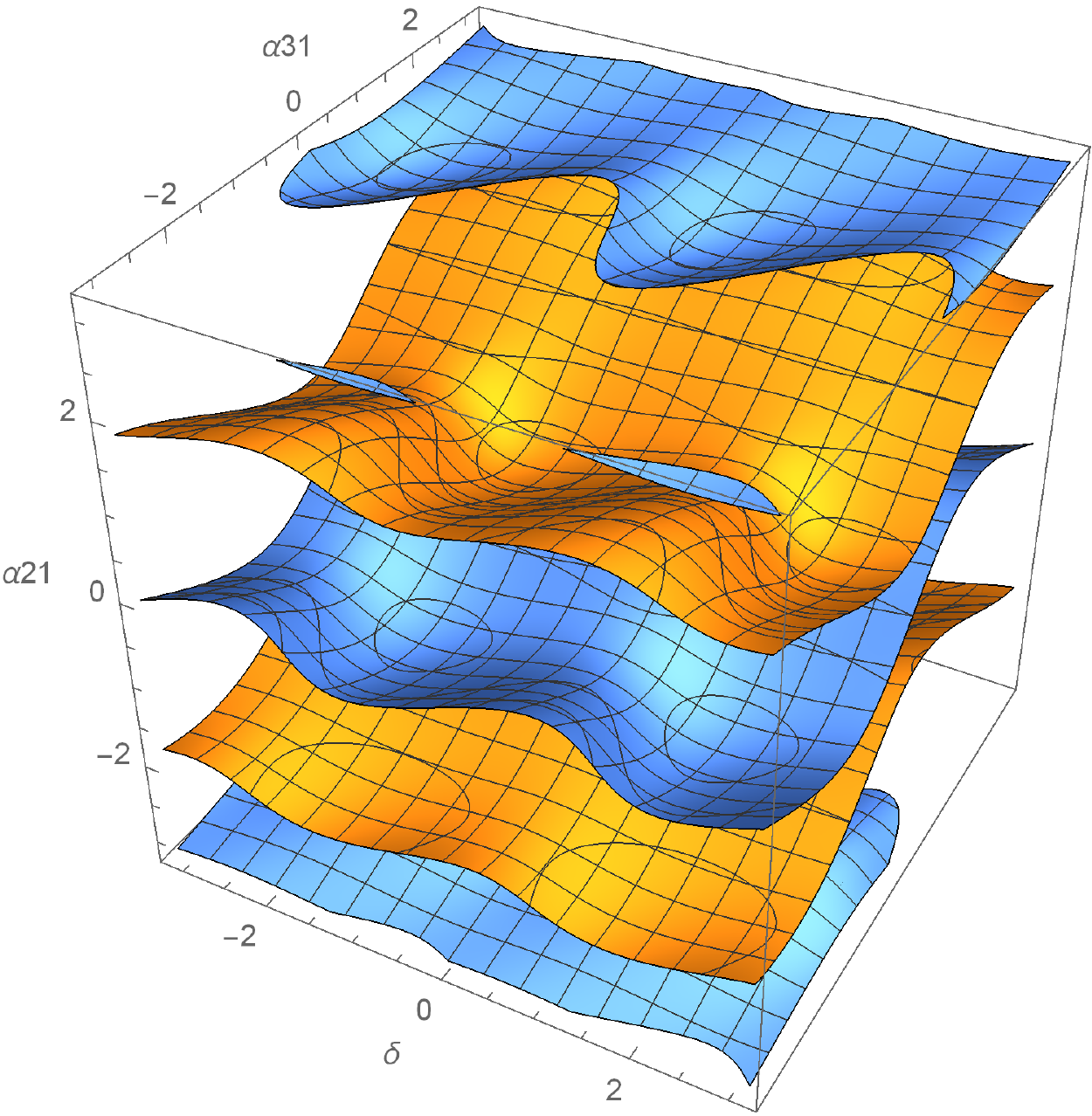}
			\caption{Contour plots on $(\delta,\alpha_{21},\alpha_{31})$ for subclass $(\rm{I\hspace{-1pt}I})$-B in IH, where $m_3$=0.046 \rm{eV}}
			\label{3D_IH_2B_10}
		\end{minipage}
	\end{tabular}
\end{figure}

\subsubsection{subclass $(\rm{I\hspace{-1pt}I})$-B in IH case}

let us explain the other case which do not lead to the experimental data, the subclass $(\rm{I\hspace{-1pt}I})$-B in IH case.
The constraints among CP phases for the $(\rm{I\hspace{-1pt}I})$-B are Eq.(\ref{Re2}) and Eq.(\ref{Im2}) by setting $\alpha$ as $e$.
The yellow and blue curved surfaces in Fig.(\ref{3D_IH_2B_01}) to (\ref{3D_IH_2B_10}) stand for Eq.(\ref{Re2}) and Eq.(\ref{Im2}) respectively.
Although two curved surfaces vary smoothly depending on the lightest neutrino mass $m_3$ from 0 \rm{eV} to 0.046 \rm{eV}, they never intersect for any intervals of $m_3$. It implies that the texture of $(\rm{I\hspace{-1pt}I})$-B do not reproduce the experimentally measured mixing angles and mass eigenvalues in IH case.

\subsubsection{$(m_{eff})_{e\tau}=0$ in NH case}
\label{sec:etau_zero}

\begin{figure}[h!]
	\begin{tabular}{ccc}
		\begin{minipage}{0.3\hsize}
			\includegraphics[width=\linewidth]{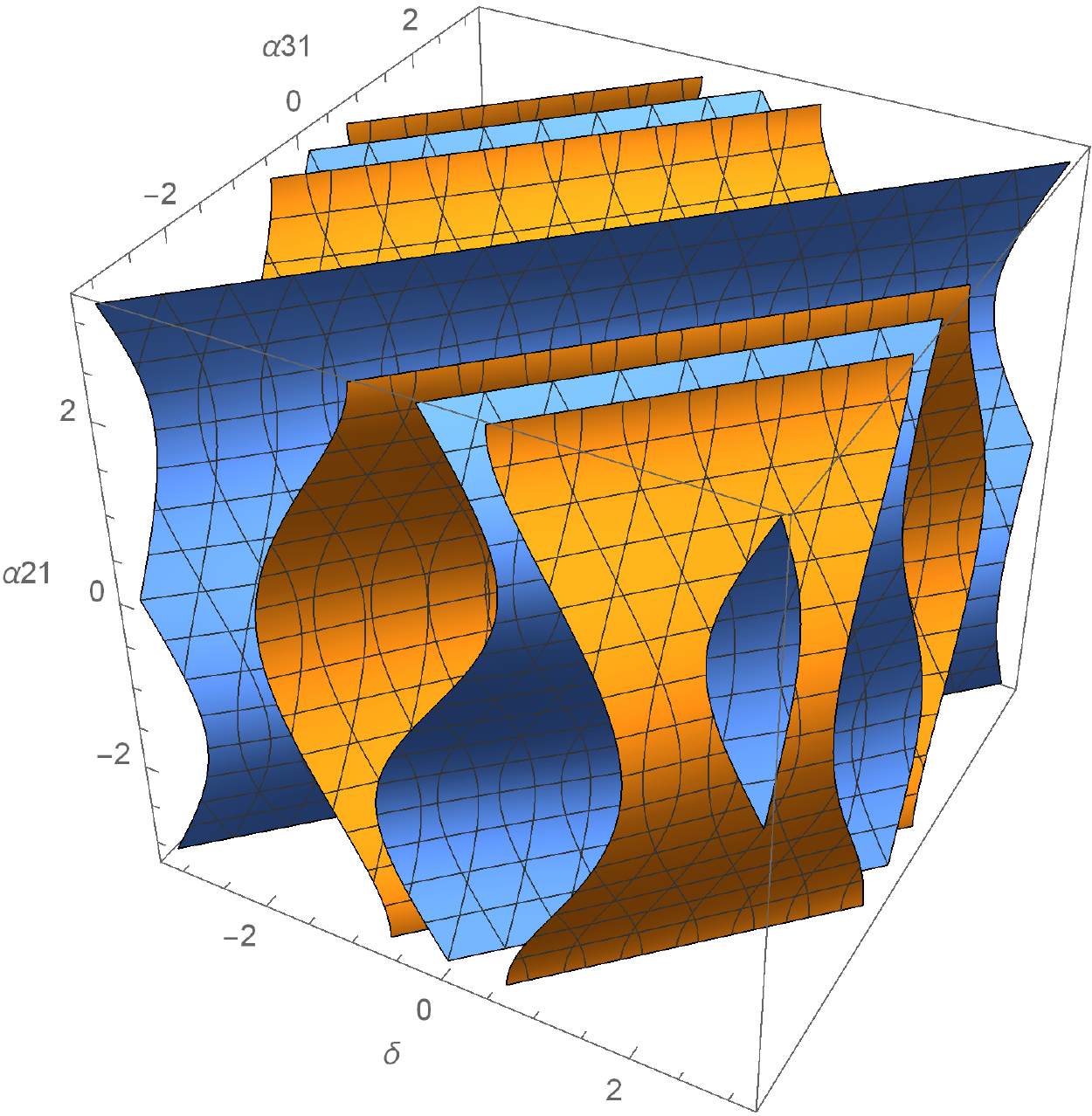}
			\caption{Contour plots on $(\delta,\alpha_{21},\alpha_{31})$ under the condition $(m_{\rm{eff}})_{e\tau}=0$ in NH, where $m_1$=0 \rm{eV}}
			\label{3D_NH_etau_01}
		\end{minipage}
		\phantom{=}
		\begin{minipage}{0.3\linewidth}
			\vspace{0cm}
			\includegraphics[width=0.975\linewidth]{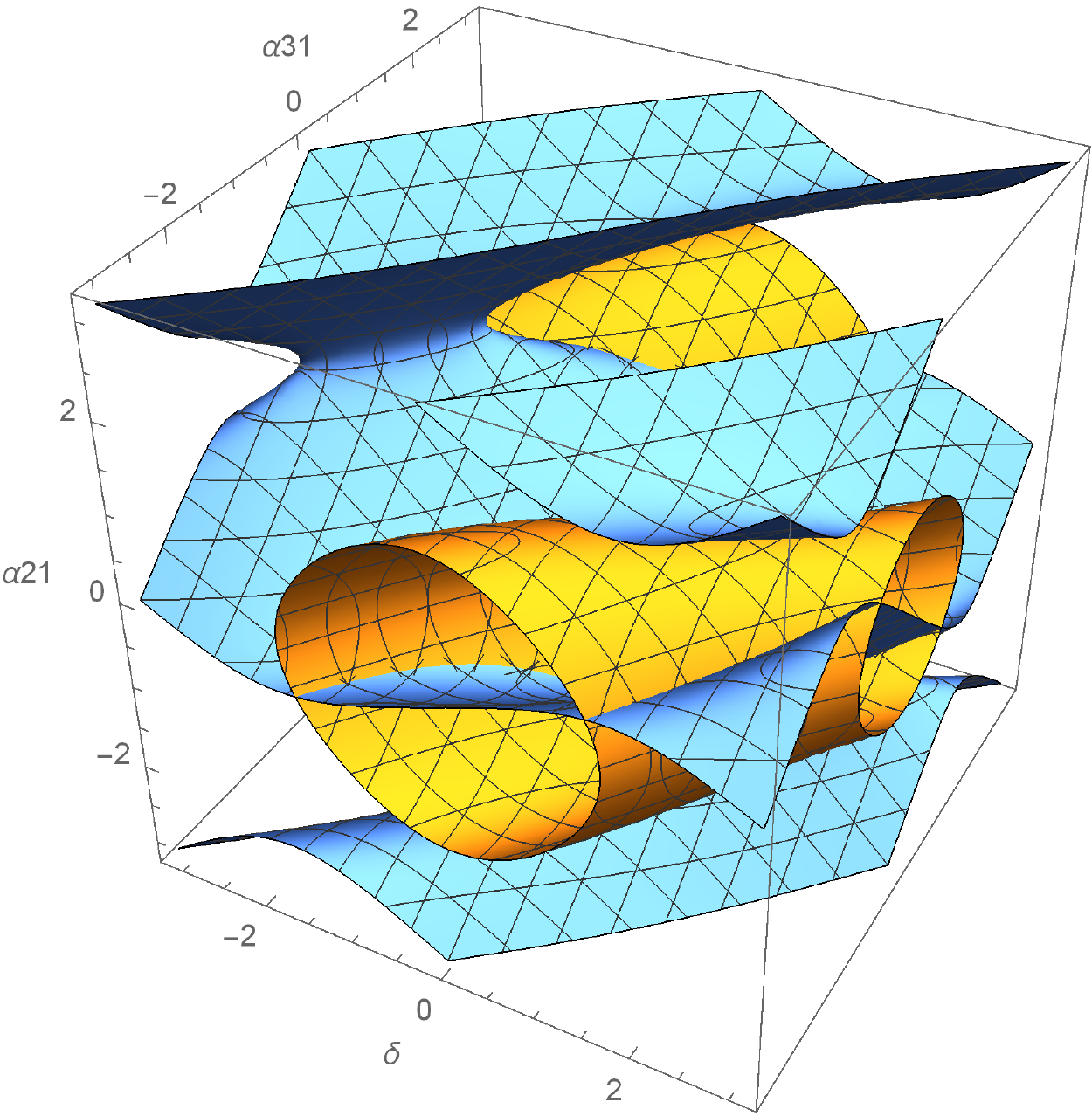}
			\caption{Contour plots on $(\delta,\alpha_{21},\alpha_{31})$ under the condition $(m_{\rm{eff}})_{e\tau}=0$ in NH, where $m_1$=0.023 [\rm{eV}]}
			\label{3D_NH_etau_05}
		\end{minipage}
		\phantom{=}
		\begin{minipage}{0.3\linewidth}
			\vspace{0cm}
			\includegraphics[width=0.975\linewidth]{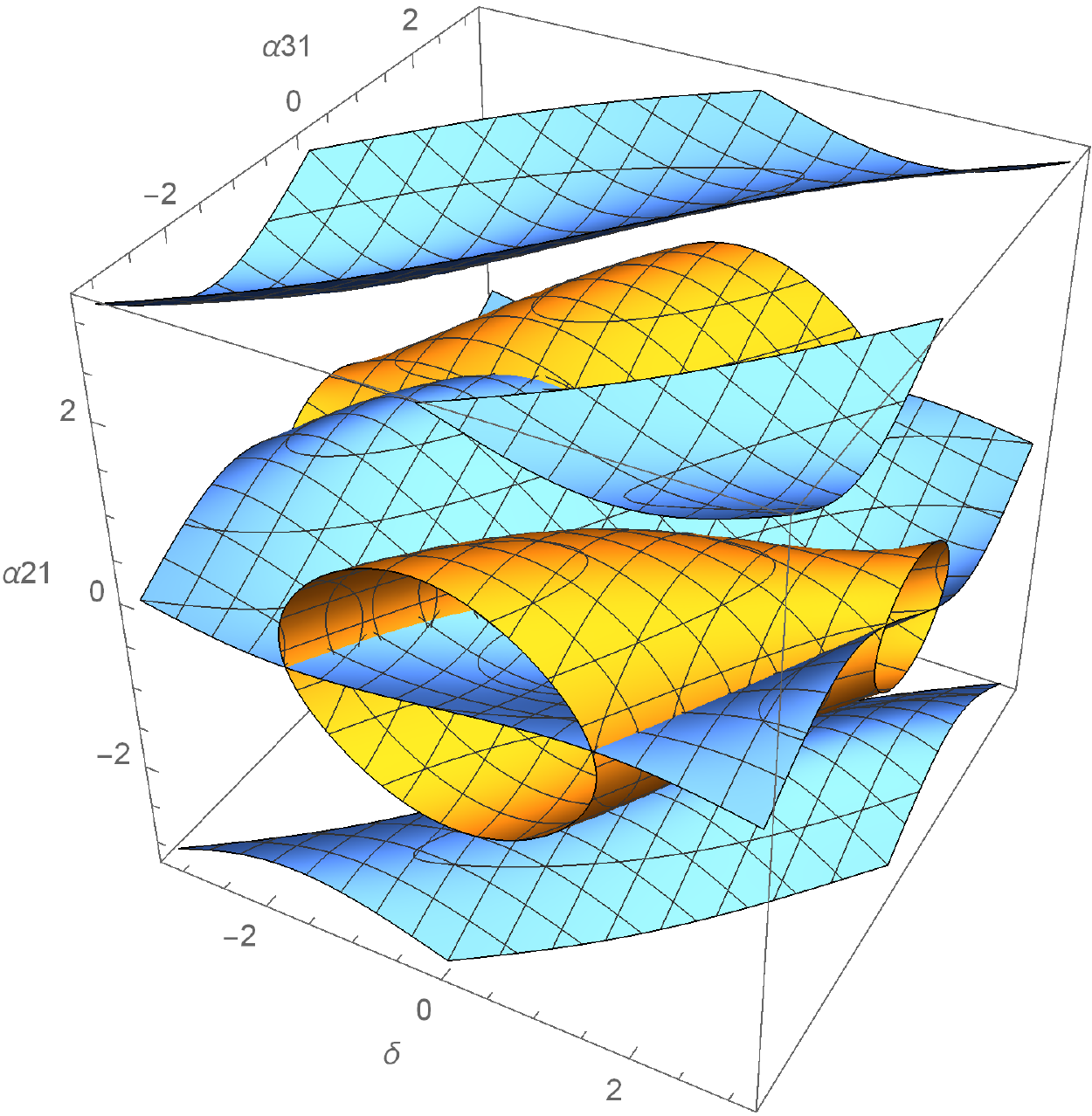}
			\caption{Contour plots on $(\delta,\alpha_{21},\alpha_{31})$ under the condition $(m_{\rm{eff}})_{e\tau}=0$ in NH, where $m_1$=0.046 \rm{eV}}
			\label{3D_NH_etau_10}
		\end{minipage}
	\end{tabular}
\end{figure}

We also mention the result in Fig.(\ref{fig.NH3ACPa21}) with constraints among $\delta$, $\alpha_{21}$ and $\alpha_{31}$ under the condition Eq.(\ref{3Dplotsf1Y2}) and Eq.(\ref{3Dplotsf2Y2}). 
In this case, there is no inconsistency between scatter plots from four-zero texture and 3D plots from one-zero effective light neutrino mass matrix. 
However, we do not find obvious relation between them in comparison with the former cases.
The yellow and blue curved surfaces in Figs.(\ref{3D_NH_etau_01})-(\ref{3D_NH_etau_10}) are described by Eq.(\ref{3Dplotsf1Y2}) and Eq.(\ref{3Dplotsf2Y2}) respectively.
The correspondence between Fig.(\ref{fig.NH3ACPa21}) and contour plots can be seen in a range of large $m_1$.

From the viewpoint of $(\delta,\alpha_{21})$ planes of Fig.(\ref{3D_NH_etau_05}) and Fig.(\ref{3D_NH_etau_10}), the nodal lines of two surfaces do not completely overlap the range of plotted points in Fig.(\ref{fig.NH3ACPa21}).
The reasons are as follows: The four-zero texture in Fig.(\ref{fig.NH3ACPa21}) is one part of the textures which produce $(m_{eff})_{e\tau}=0$. There is no one-to-one correspondence between the results from Fig.(\ref{fig.NH3ACPa21}) in numerical analysis and from Figs.(\ref{3D_NH_etau_01})-(\ref{3D_NH_etau_10}).  Besides, only the central values of mixing angles and mass squared differences are chosen to draw the 3-dimensional plots, while these values in the scattered plots are taken within $3$ standard deviations from the mean of the experimental values.

\section{Dscussions and Summary}\label{sec:con}
We have discussed the effective mass matrix for left-handed Majorana neutrinos $m_\text{eff}$.  It is expressed with nine phenomenological parameters, i.e., two mass squared differences, the lightest neutrino mass, three lepton mixing angles, one Dirac and two Majorana phases. In this paper, we have investigated the 
four-zero texture model for the Dirac neutrino mass matrix. After using Type-I seesaw mechanism, we have obtained 
$m_\text{eff}$ which includes seven model parameters 
($\theta _1$, $\theta _2$, $\phi _1$, $\phi _2$, $X_1$, $X_2$, and $X_3$).
We have classified $_9C_4=126$ different textures for Dirac neutrino mass matrix into seven classes.
We have investigated three classes (I), $(\rm{I\hspace{-1pt}I})$ and $(\rm{I\hspace{-1pt}I\hspace{-1pt}I})$ which lead to non-zero lightest neutrino mass, non-degenerate light neutrino mass spectrum,  and non-zero Jarlskog invariant.  

In our numerical analysis, we have taken the model parameters which reproduce two mass squared differences and three mixing angles within $3$ standard deviations from the mean of the experimental values. Since the number of model parameters is reduced to seven, we can predict the correlations among CP violating phases after assuming
the neutrino mass hierarchy and the value of the lightest neutrino mass. 
For each class, we do not have to repeat the procedure to investigate all textures according to the relation of the
configurations among some different textures. 
The correlations between the Dirac CP violating phase and one of the Majorana phases are plotted. 
Some textures show that a CP violating phase is restricted in narrow range, 
others show the clear correlation between two CP violating phases.

We have also made analysis from the view point of the effective Majorana neutrino mass matrix $m_\text{eff}$.
Some textures of the Dirac mass matrix produce $m_\text{eff}$ which have one zero off-diagonal element.
One of the off-diagonal elements of the Majorana mass matrix should vanish so that we have obtained two constraints. 
The other textures, which lead to nine non-vanishing elements of Majorana mass matrix, 
also produce two constraints. This is because we have assumed four-zero textures on the Dirac mass matrix.
We found the hidden relations among CP violating phases after assuming the value of the lightest neutrino mass.
These hidden relations correspond to the four-zero texture for the Dirac neutrino mass matrix. 
In particular, the subclasses which do not produce any scattered plots in numerical analysis are explained by means of 3D plot among CP violating phases. 
Such subclasses made no nodal lines of surfaces described by the conditions.

\vspace{5mm}
\noindent
{\bf Acknowledgement}
We would like to thank Prof. M.Tanimoto for useful comments.
The work of T.M. and Y.S. are supported by JSPS KAKENHI Grant Number JP17K05418 and
the work of Y.S. is also supported  by Fujyukai Foundation. 
The work of H.U. is supported in part by the Ministry of Science and Technology of R.O.C. under Grant No. MOST 107-2811-M-001-025.
\newpage 
\appendix 


\newpage


\end{document}